\documentclass[%
reprint,
superscriptaddress,
 amsmath,amssymb,
 aps,
prb,
]{revtex4-2}

\usepackage{graphicx}
\usepackage{dcolumn}
\usepackage{bm}
\usepackage{hyperref}
\usepackage{placeins}
\usepackage{xcolor}

\newcommand{\G}{\mathcal{G}}

\begin{document}

\preprint{APS/123-QED}
\title{Hamiltonian learning for spin-spiral moir\'e magnets from electronic magnetotransport}

\author{Fedor Nigmatulin}
\email{fedor.nigmatulin@aalto.fi}
\affiliation{Department of Electronics and Nanoengineering, Aalto University, FI-00076 Aalto, Finland}

\author{Greta Lupi}
\affiliation{Department of Applied Physics, Aalto University, FI-00076 Aalto, Finland}

\author{Jose L. Lado}
\affiliation{Department of Applied Physics, Aalto University, FI-00076 Aalto, Finland}

\author{Zhipei Sun}
\affiliation{Department of Electronics and Nanoengineering, Aalto University, FI-00076 Aalto, Finland}

\date{\today}

\begin{abstract}
Two-dimensional noncollinear magnetic states, such as spin-spiral magnets, offer an excellent platform for investigating fundamental phenomena, with potential for advancing stray-field-free spintronics. However, detection and characterization of noncollinear magnetic states in two-dimensional systems remain challenging, motivating the development of alternative probing methods. Here, we present a methodology for extracting the spin-spiral $\mathbf{q}$ vector from lateral electronic transport measurements. Our approach leverages the magnetic field and bias dependence of the conductance to train a supervised machine learning algorithm, which enables us to extract the $\mathbf{q}$ vectors of arbitrary spin-spiral magnets. We demonstrate that this methodology is robust to the presence of impurities in the system
and noise in the conductance data. Our findings show that the conductance pattern reveals a complex dependence on the $\mathbf{q}$ vector of the spin spiral, providing a new strategy to learn magnetic structures directly from transport experiments. 
\end{abstract}

\maketitle

\section{Introduction}
\label{sec:introduction}

Noncollinear magnetic structures offer a rich playground for both fundamental condensed matter physics and future spintronics applications. Two-dimensional (2D) noncollinear magnetic systems are of particular interest due to their promising integrability as building blocks of van der Waals (vdW) heterostructures and their accessible tunability, for example, via electrostatic gating \cite{mak2019}. However, the detection and characterization of noncollinear magnetization in 2D systems remain elusive \cite{Song2025, Wang2024, Miao2025, Nigmatulin2025, Amini2024}. Hamiltonian learning \cite{Wang2017, Valenti2022, Koch2025, Barey2019, Gebhart2023, Lupi2025, vandriel2024, Karjalainen2023, Karjalainen2025, Koch2022, Koch2023, Valenti2022, Che2021, Simard2025} applied to experimental data can provide an efficient strategy for tackling this problem. This approach enables the extraction of Hamiltonian parameters of magnetic systems and the distinction between different magnetic phases \cite{Wang2020, SALCEDOGALLO2020, Iakovlev2018, Feng2024}. Various experimental strategies have been proposed for learning the Hamiltonian of quantum materials, including inelastic spectroscopy with scanning tunneling microscopy (STM) \cite{Koch2025, Karjalainen2023, Koch2022,lupi2026} and mesoscopic transport \cite{Koch2023,vandriel2024,hernandes2026}. In particular, electronic transport experiments offer a potential strategy to learn complex phenomena in 2D materials.

\begin{figure}[t!]
    \centering
    \includegraphics[width = 8.6cm]{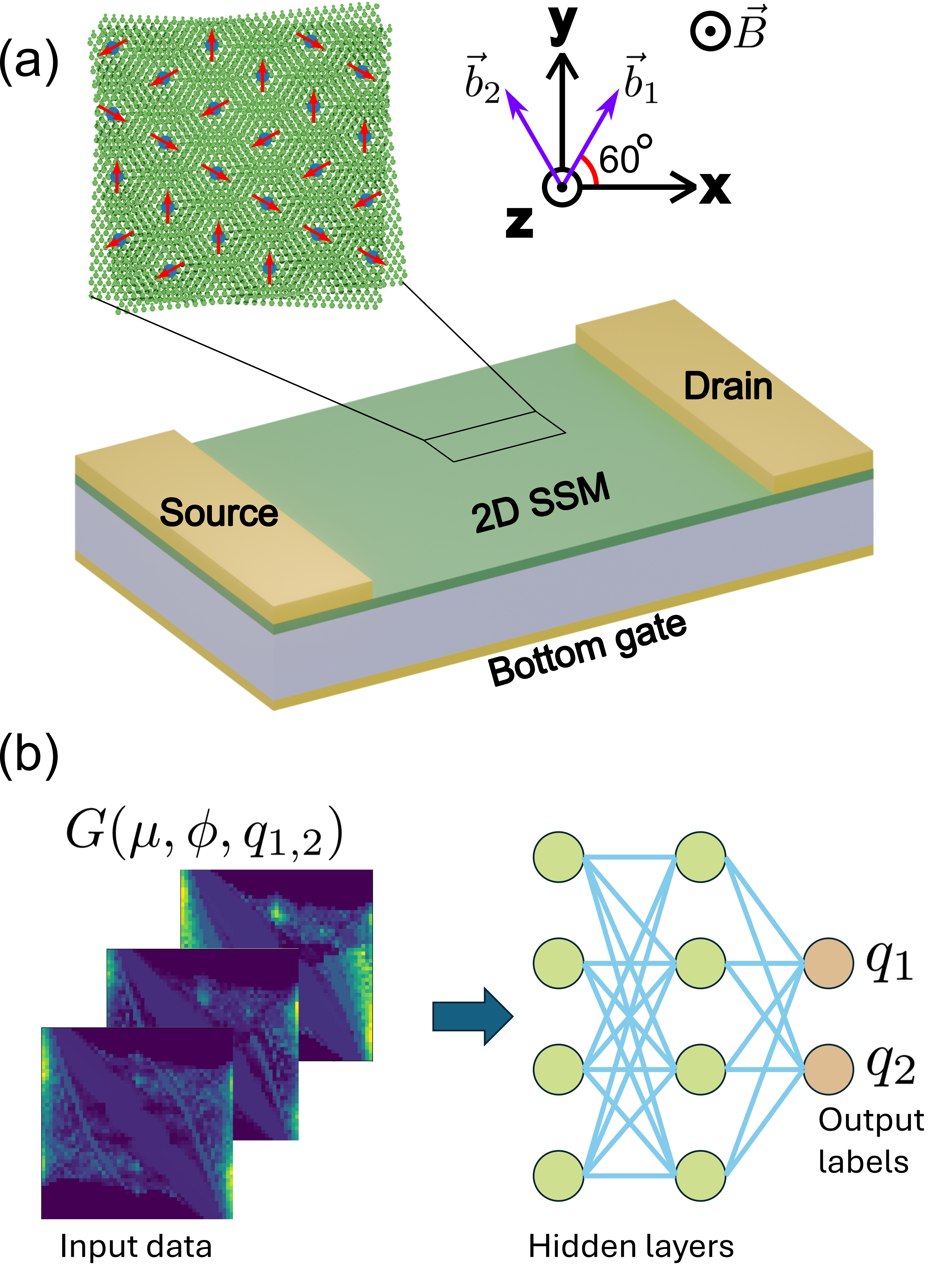} 
    \caption{(a) Schematic of an electrical gated device to probe noncollinear magnetic order in a twisted vdW structure (see the inset). Red arrows show local magnetic moments localized on the sites of the triangular moir\'e superlattice. $\mathbf{b}_{1,2}$ are reciprocal lattice basis vectors and $\vec{B} = B \mathbf{e}_z$ is a perpendicular external magnetic field. (b) Schematic representation of the Hamiltonian learning approach employed for the extraction of the $\mathbf{q} = (q_1, q_2)$ of the SSM by
    leveraging the conductance $G(\mu, \phi, q_{1,2})$.}
    \label{fig:fig1}
\end{figure}

Here, we demonstrate a machine learning (ML) methodology that combines magnetic field‑ and bias‑dependent transport to infer the spin‑spiral order of a 2D magnet. Our algorithm enables the extraction of the spin-spiral $\textbf{q}$ vector, which is the parameter that defines the period and propagation direction of the spin spiral. To observe signatures of spin-spiral magnets (SSMs) contained in the electronic transport data, we consider Hofstadter butterfly-type conductance \cite{Hofstadter1976} as a function of electron doping and magnetic flux. We leverage the impact of the spin spiral on the Hofstadter pattern, focusing on an effective triangular superlattice of a twisted vdW structure, showing that such a transport signature allows reconstruction of the
original spin spiral.

Our paper is organized as follows. In Sec.~\ref{sec:model}, we describe a model of the electrical gated device with the 2D spin-spiral magnetization in the channel. The discussion of the conductance simulations of the considered system is presented in Sec.~\ref{sec:electronic_transport}. In Sec.~\ref{sec:neural_network_architecture}, we introduce the Hamiltonian learning approach and discuss the prediction results for the $\mathbf{q}$ vector. In Sec.~\ref{sec:noise_study}, we discuss the inclusion of noise in the conductance data to emulate a real experiment. We close with our conclusions, given in Sec.~\ref{sec:conclusions}.

\section{Model}
\label{sec:model}

We will focus on a system consisting of a tunable 2D electron gas, whose electronic transport is used to probe a spin spiral placed in its proximity. Such a system can be realized in a twisted transition metal dichalcogenide (TMD) multilayer heterostructure composed of two twisted bilayers. The first twisted bilayer is tuned to the strongly correlated
Mott regime, so that it gives rise to a spin-spiral ordered state \cite{Wu2018, regan2020}. The second twisted bilayer is set to a narrow band metal limit, realizing an electron gas with an external gate-tunable electron density \cite{foutty2025, kometter2023, Zhao2025, Foutty2024}, which is the sensing layer where electronic transport is measured.
Figure~\ref{fig:fig1}(a) shows the schematic of the proposed electrical gated device to probe 2D noncollinear magnetic order. The fundamental quantity we use is the conductance provided by the source and drain electrodes, as a function of the electrostatic gate, which effectively controls the electronic density of the tunable 2D electron gas, and an external magnetic field. Our strategy relies on leveraging the Hofstadter regime of the metallic moir\'e superlattice, which can be achieved with moderate magnetic fields thanks to the enlarged moir\'e superlattice constant \cite{Hunt2013, Dean2013, ponomarenko2013, foutty2025}.
We exploit the modification of the Hofstadter butterfly pattern caused by the exchange coupling with the 2D spin spiral. These modifications reflect the underlying spin spiral pattern and enable us to extract the desired $\mathbf{q}$ vector. The model example of 2D SSM in the twisted 2D structure is shown in the inset of Fig.~\ref{fig:fig1}(a).
The proposed device can be made with the means of modern fabrication techniques \cite{mak2022, He2021, lau2022}. Deterministic transfer method, for example, can be used for mechanical stacking of the 2D vdW structure \cite{Hao-Wei2021}, and electron beam lithography followed by physical vapor deposition can provide electrodes \cite{Xu2016, jain2019}. 

The effective Hamiltonian projected onto the low-energy Wannier states of the metallic twisted moir\'e system takes the form
\begin{equation*}
H = t\sum_{\left\langle \alpha\beta \right\rangle s}\left(e^{i \phi_{\alpha\beta}} c_{\alpha s}^\dagger c_{\beta s}+h.c.\right) + \mu \sum_{\alpha s} c_{\alpha s}^\dagger c_{\alpha s} +
\end{equation*}

\begin{equation}
+ \sum_{\alpha ss'}\left(\mathbf{J}_\alpha(\mathbf{q})\cdot \vec{\sigma}\right)_{ss'} c_{\alpha s}^\dagger c_{\alpha s'} + \sum_{\alpha s}W_\alpha c_{\alpha s}^\dagger c_{\alpha s},
\label{Hamiltonian}
\end{equation}
where $t_{\alpha \beta }$ are the nearest neighbor hoppings between Wannier moir\'e orbitals, $c_{\alpha s}^\dagger$, $c_{\alpha s}$ are the creation and annihilation operators, respectively, for an electron with spin projection $s$ at the site $\alpha$ of the triangular moir\'e superlattice,  $\phi_{\alpha\beta} = \int_{\vec r_\alpha}^{\vec r_\beta} \vec A \cdot d \vec l$ is the magnetic phase created by the vector potential $\vec A$, with $\vec B = \nabla \times \vec A$, $\mu$ is the chemical potential, and $\vec{\sigma}$ is a vector of Pauli matrices, representing the electron spin degree of freedom. The nonmagnetic disorder in the system is included by the Wannier on-site energies $W_\alpha \in [-W/2, W/2]$, where $W$ denotes the disorder strength. In a real crystal, the disorder stems primarily from twist disorder of the moir\'e superlattice, or underlying atomic defects within the layers. The local exchange field $\mathbf{J}_\alpha$ created by the 2D spin-spiral magnetization onto the electron gas takes the form
\begin{equation}
\mathbf{J}_\alpha = J \left(\sin \mathbf{q} (\mathbf{r}_\alpha - \mathbf{r}_0),\, \cos{\mathbf{q}  (\mathbf{r}_\alpha - \mathbf{r}_0)}, \, 0 \right),
\label{magnetization}
\end{equation} 
where $J$ is the exchange coupling and $\mathbf{r}_\alpha$ is the coordinate of the lattice site $\alpha$. The $\mathbf{q}$ vector is defined in terms of the reciprocal lattice basis vectors $\mathbf{b}_{j = 1,2}$ (purple arrows in Fig.~\ref{fig:fig1}(a)): $\mathbf{q} = q_1 \mathbf{b}_1 + q_2 \mathbf{b}_2$. We will quantify $q_{j=1,2}$ in units of $\frac{2\pi}{d}$, where $d$ is the lattice constant of the triangular superlattice. Thus, in this notation, $\mathbf q = (0,0)$ corresponds to the ferromagnetic order, and finite $\mathbf q$ corresponds to different types of spiral order.

\begin{figure}[t]
    \centering
    \includegraphics[width = 8.6cm]{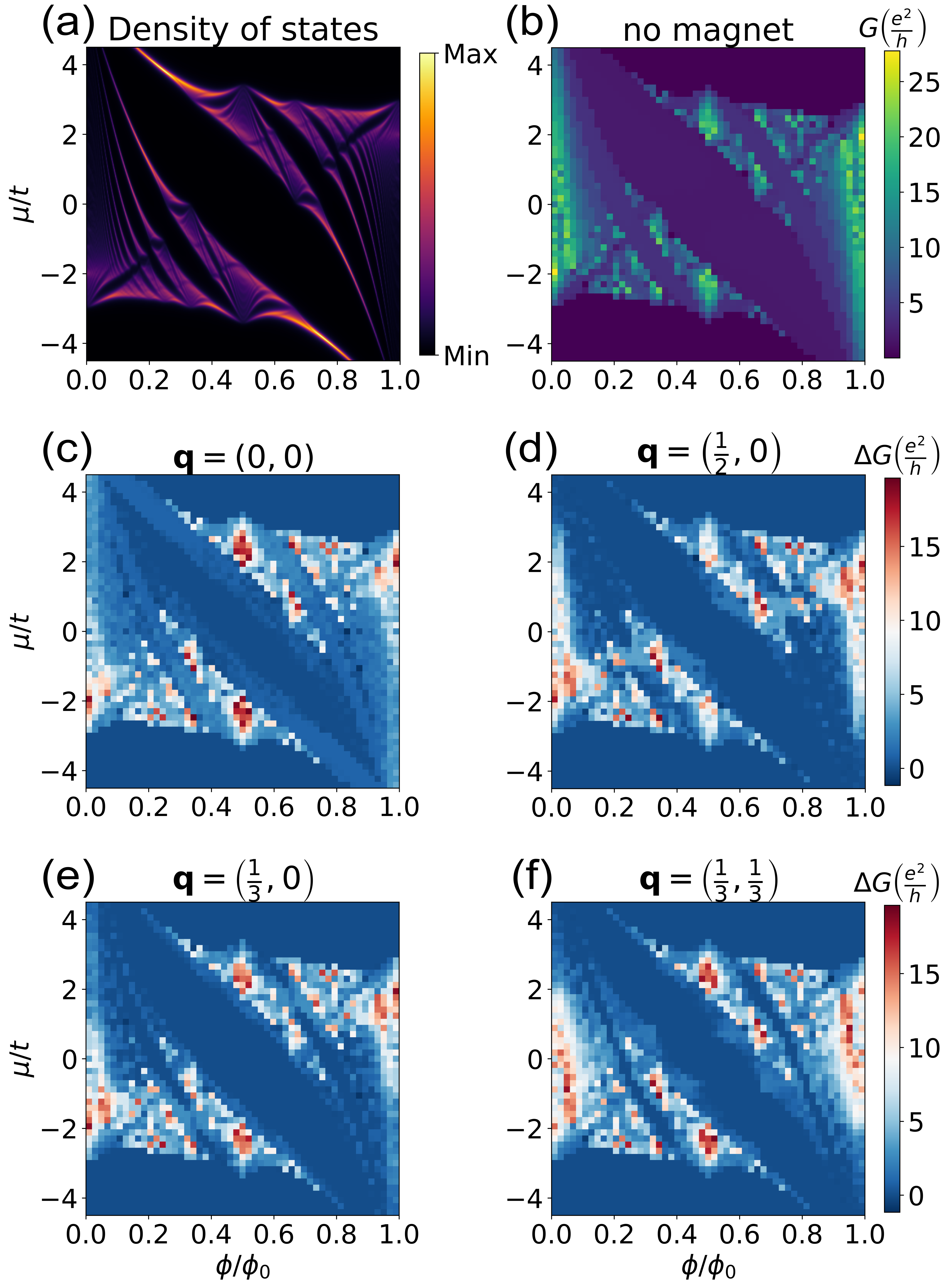} 
    \caption{(a) Density of states of the Hofstadter butterfly for a triangular-lattice nanoribbon as a function of the chemical potential $\mu$ and normalised magnetic flux $\phi/\phi_0$. (b) Conductance without a magnetic exchange proximity. (c)-(f) Exchange proximity-induced conductance change $\Delta G$ for different special values of $\mathbf{q}$ calculated using \eqref{Landauer_conductance} for $J = t$ and $W = 0.2t$.}
    \label{fig:fig2}
\end{figure}

Figure~\ref{fig:fig2}(a) shows the density of states of the pristine moir\'e electron gas
as a function of the chemical potential in units of $t$ and the magnetic flux normalized by the magnetic flux quantum $\phi_0 = h/e$, presenting a fractal Hofstadter butterfly pattern. While this is solely the spectrum of the system, as discussed below, the conductance in moderate junctions, with a length smaller than the magnetic length, will directly reflect such a Hofstadter butterfly pattern. In the following, we will focus on systems in which the coherence length is larger than the channel length, so that the current is dominated by
ballistic phase-coherent transport. 

\section{Electronic transport through the 2D SSM}
\label{sec:electronic_transport}
In the ballistic phase-coherent limit, the current through the moir\'e metallic channel can be found from the Landauer formula as

\begin{equation}
    I = \frac{2e}{h} \int dE (f(E, \mu_S)-f(E, \mu_D))\mathrm{Tr}(\G^\dagger \Gamma_D \G \Gamma_S),
    \label{Landauer}
\end{equation}
where $f(E, \mu)$ is the Fermi-Dirac distribution; $\G$ is the Green's function of the channel,
$\Gamma_{S, D} = i (\Sigma_{S, D} - \Sigma_{S, D}^\dagger)$, where $\Sigma_{S, D}$ are the lead self-energies, and $\mu_{S, D}$ are the chemical potentials of the source and drain, respectively. In the linear response limit of a small bias voltage and at low temperatures, the conductance derived from Eq.~\eqref{Landauer} takes the following form \cite{bruus2004many}

\begin{equation}
    G = \frac{2e^2}{h} \mathrm{Tr}(\G^\dagger \Gamma_D \G \Gamma_S).
    \label{Landauer_conductance}
\end{equation}
We used Eq.~\eqref{Landauer_conductance} and the nonequilibrium Green's function formalism to calculate the conductance of the considered system as a function of the chemical potential and the external magnetic field \cite{pyqula}. Figure~\ref{fig:fig2}(b) shows the calculated conductance without SSM exchange proximity ($J = 0$) that exhibits the Hofstadter butterfly pattern similar to the density of states in Fig.~\ref{fig:fig2}(a). It is worth noting that the presence of a magnetic field gives rise to chiral quantum Hall edge states, revealed by the nonzero conductance within the bulk gap, as seen in Fig.~\ref{fig:fig2}(b) \cite{Klitzing1980, Halperin1982}. We study how the SSMs modify this Hofstadter pattern for different values of $\mathbf{q}$. Examples of the simulated conductance change for different $\mathbf{q}$ are shown in Fig.~\ref{fig:fig2}(c)-(f). These conductance maps allow us to directly extract the spin-spiral \textbf{q} vector as described below.

\begin{figure}[t]
    \centering
    \includegraphics[width = 8.6cm]{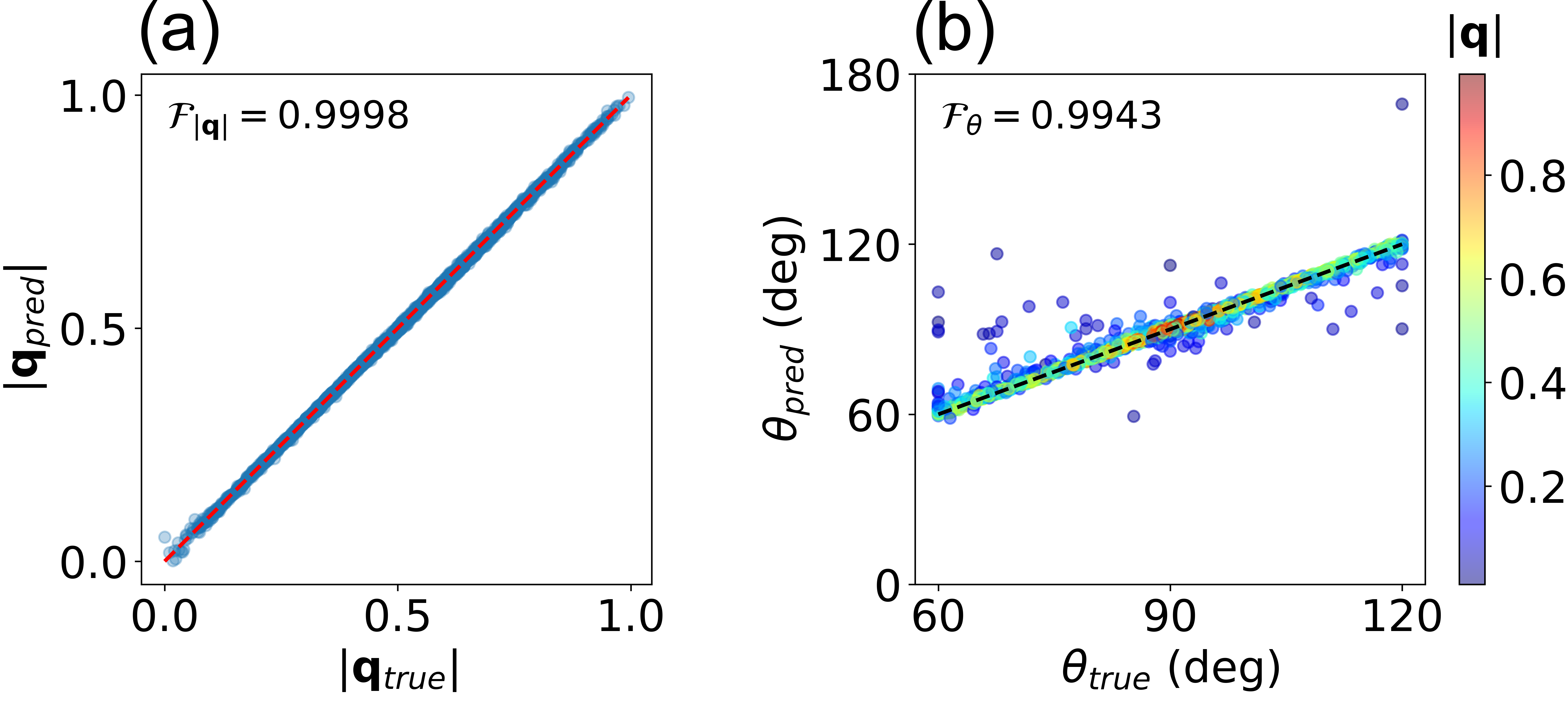} 
    \caption{Prediction results for $|\mathbf{q}|$ (a) and $\theta$ (b) for previously unseen inputs with the number of PCs $N_{PCA} = 500$, $J = t$, and $W = 0.2t$.
    }
    \label{fig:fig3}
\end{figure}

\section{Hamiltonian-learning spin-spiral magnets}
\label{sec:neural_network_architecture}
The results of the conductance simulations reveal a complex dependence on the $\mathbf{q}$ vector, which represents the fundamental input of the ML algorithm. For this purpose, we generated a conductance dataset with 10000 samples, analogous to those shown in Fig.~\ref{fig:fig2}(b)-(f), with the $q_1$ and $q_2$ components spanning from 0 to 1/2. 
We trained a supervised neural network (NN) with the simulated conductance dataset to
extract the $\mathbf{q}$ vector directly from the conductance maps. The initial dataset was labeled with randomly distributed $q_{1,2}$ components and split into training (6800 samples), validation (1200 samples), and testing (2000 samples) subsets. Figure~\ref{fig:fig1}(b) shows the schematic representation of the NN's architecture with two hidden layers of 100 neurons each and two outputs that give $q_1$ and $q_2$ values. Principal component analysis (PCA) was applied to the input data to reduce dimensionality and improve the robustness to noise. The Appendix describes the PCA and provides additional details on the training process. 

We characterize the accuracy of the spin spiral prediction using fidelity parameters, separated into polar angle $\theta$ and magnitude $|\mathbf{q}|$ components. We define angular fidelity as

\begin{equation}
    \mathcal{F}_{\theta} =
\Bigg\langle \cos \left(2\left({\theta_{pred}-\theta_{true}}\right)\right) \Bigg\rangle,
\label{fidelity_angle}
\end{equation}
ensuring invariance under $\theta \rightarrow \theta + n \pi$ where $n \in \mathbb{Z}$.
It should be noted that since our device has an inversion symmetry, $\mathbf q$ and $-\mathbf q$ are indistinguishable from a transport point of view. The $|\mathbf q|$-fidelity, is computed as \cite{Lupi2025, Khosravian2024,lupi2026} given by

\begin{equation}
    \mathcal{F}_{|\mathbf{q}|} =
\frac{\langle |\mathbf{q}^{\mathrm{pred}}| |\mathbf{q}^{\mathrm{true}}| \rangle -
\langle |\mathbf{q}^{\mathrm{pred}}| \rangle \langle |\mathbf{q}^{\mathrm{true}}| \rangle}{\sqrt{\operatorname{var}\left(|\mathbf{q^{\mathrm{pred}}}|\right)\, \operatorname{var}\left(|\mathbf{q}^{\mathrm{true}}|\right)}},
\label{fidelity_length}
\end{equation}
where $\mathbf{q}^{\mathrm{pred}}$ and $\mathbf{q}^{\mathrm{true}}$ are predicted and true vectors, respectively, and $\operatorname{var}(\cdot)$ stands for variance.

In these fidelity notations, $\mathcal{F}_{|\mathbf{q}|,\theta}= 0$ indicates an absence of prediction, and $\mathcal{F}_{|\mathbf{q}|,\theta}=1$ corresponds to a perfect prediction. Figures~\ref{fig:fig3}(a) and (b) show that the NN effectively predicts unknown $\mathbf{q} = (q_1, q_2) = |\mathbf{q}|e^{i\theta}$, obtained with 500 principal components (PCs). After training the NN,
the fidelities reach $\mathcal{F}_{|\mathbf{q}|}=0.9998$ and $\mathcal{F}_{\theta}=0.9943$. 

\begin{figure}[t!]
    \centering
    \includegraphics[width = 8.6cm]{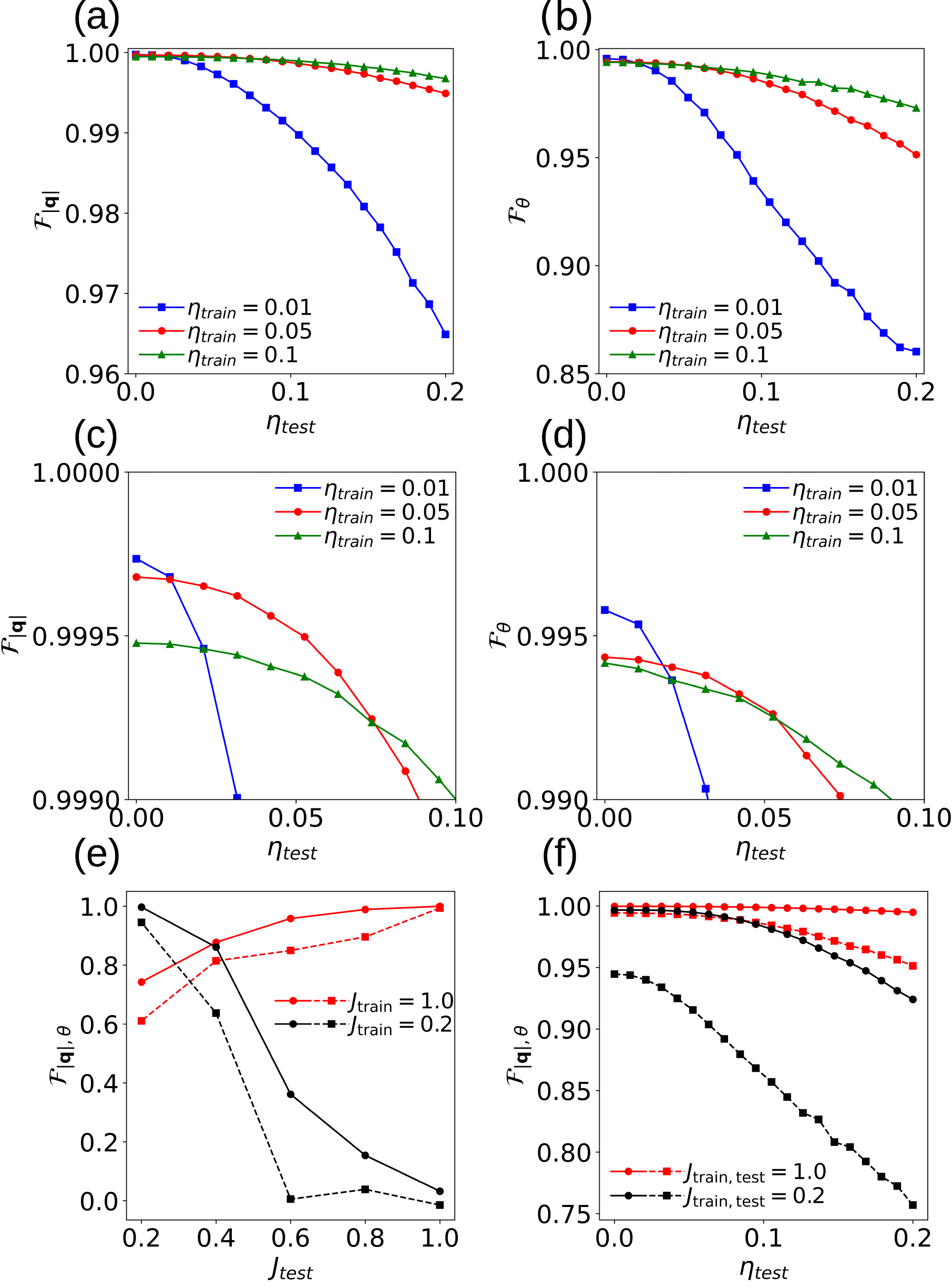} 
    \caption{(a)-(d) The dependence of the fidelity as a function of the noise strength $\eta_{test}$ in the conductance for different noise strengths used in the training $\eta_{train}=0.01,\, 0.05,\, 0.1$. (e) The dependence of the fidelity vs exchange coupling $J_{test}$ in testing data for different exchange couplings used in the training $J_{train}=0.2t$ and $t$. (f) The fidelity vs noise dependence for different exchange couplings $J_{train}= J_{test} = 0.2t$ and $t$. Solid and dashed lines in (e)-(f) show the $|\mathbf{q}|$-fidelity $\cal F_{|\bf{q}|}$ and angular fidelity $\cal F_{\theta}$, respectively. 
}
    \label{fig:fig4}
\end{figure}

\begin{figure}[t!]
    \centering
    \includegraphics[width=8.6cm]{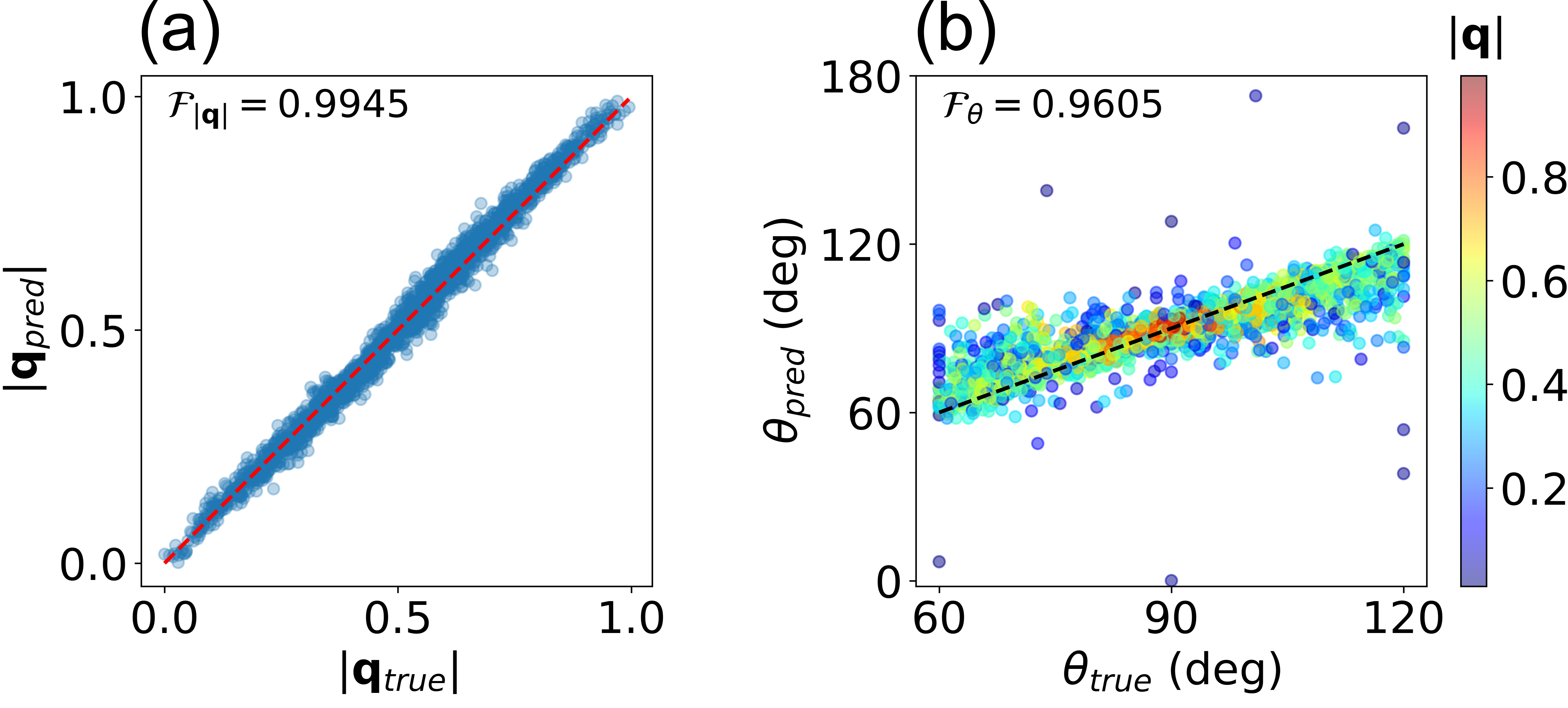}
    \caption{Prediction results for $|\mathbf{q}|$ (a) and $\theta$ (b) obtained with the NN trained and tested with the noisy data with $\eta_{train} = 0.05$ and $\eta_{test} = 0.05$; $N_{PCA} = 500$, $J = t$, and $W = 0.2t$.
    }
    \label{fig:fig5}
\end{figure}

\section{Resilience to noise}
\label{sec:noise_study}
The real experimental data are inevitably affected by noise. Robustness to noise is a key condition for applying this methodology to realistic experimental measurements. We use the angular and $|\mathbf{q}|$-fidelity defined by Eqs.~\eqref{fidelity_angle} and \eqref{fidelity_length}, respectively, to analyze the noise resilience of the prediction results for the different parameters of the NN and the Hamiltonian~\eqref{Hamiltonian}. We introduce scaling bounded noise in the conductance as follows
\begin{equation}
G_{\eta} = G\left (1 + \frac{\eta}{2}\right )^2,
\end{equation}
where $G$ is a noise-free conductance, $\eta\in [-\eta_0, \eta_0]$, and $\eta_0$ controls the noise strength. Figures~\ref{fig:fig4}(a)-(d), (f) show the fidelity decay with an increase in $\eta_{test}$ of the noise added to the transport data. We quantified the performance of our model trained with noisy data with noise strengths $\eta_{train} = 0.01, \,0.05,\, 0.1$ (see Fig.~\ref{fig:fig4}(a)-(d)). The addition of noise to the training data leads to an increase in the dispersion in the prediction results and a decrease in fidelity, as seen in Figs.~\ref{fig:fig5}(a)-(b). It is worth noting that while a higher noise level in the training data results in worse fidelity for the noise-free testing data, it starts showing better performance for the significantly noised signal. The latter can be seen, for example, in Figs.~\ref{fig:fig4}(c)-(d) by the crossing of the red curves with $\eta_{train} = 0.1$ with the blue ($\eta_{train} = 0.01$) and green ($\eta_{train} = 0.05$) fidelity curves around $\eta_{test} = 0.01-0.02$ and $0.05-0.07$, respectively.

\begin{figure}[t!]
    \centering
    \includegraphics[width=8.6cm]{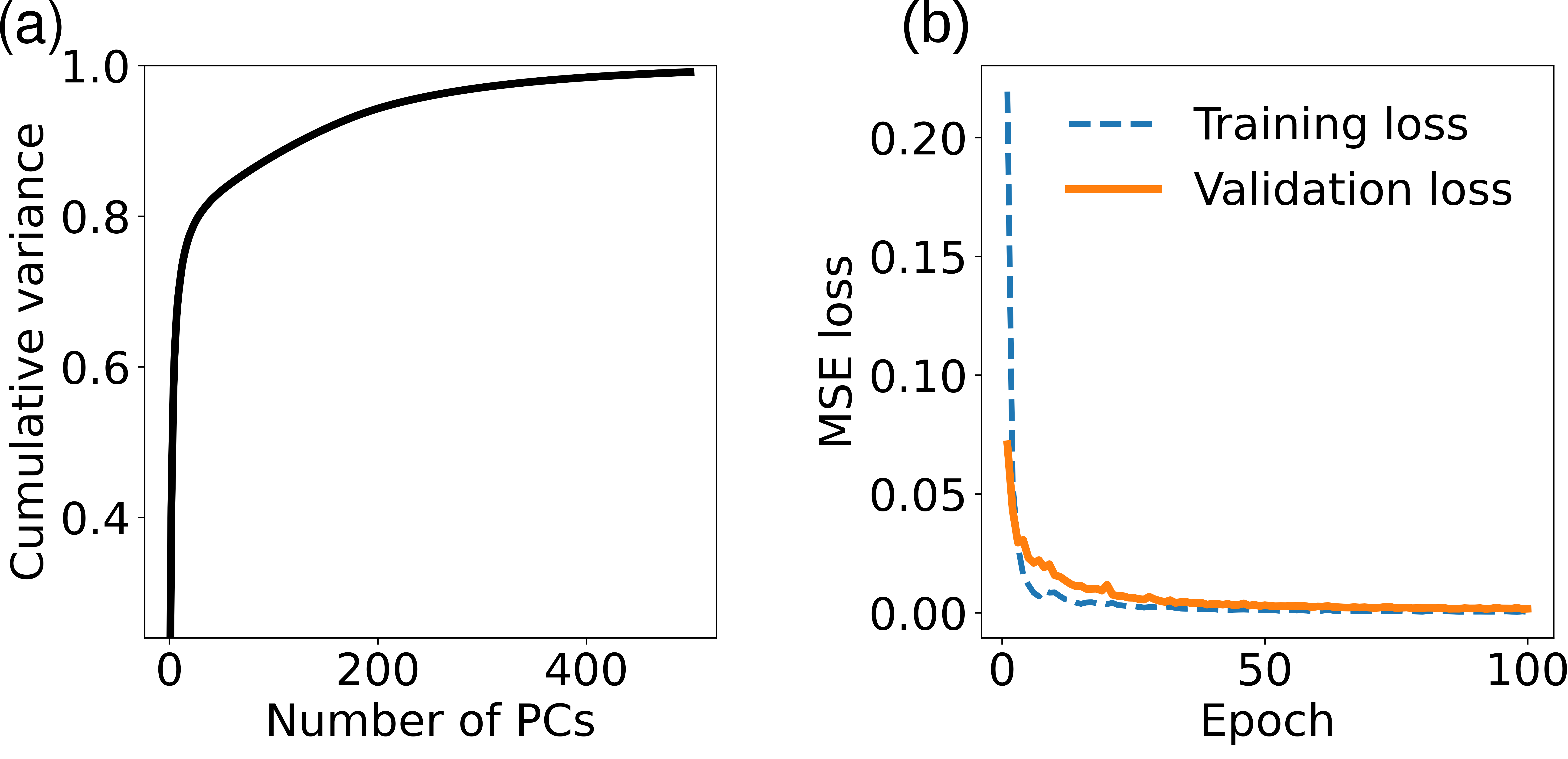}
    \caption{(a) Cumulative variance as a function of the number of PCs. (b) The MSE loss as a function of the epoch for training (blue dashed curve) and validation datasets (orange solid curve).}
    \label{fig:fig6}
\end{figure}

A crucial parameter affecting the fidelity of our algorithm is the exchange coupling $J$. We tested the performance of the ML algorithm for different exchange couplings $J_{train, test}$ in the testing and training datasets as shown in Fig.~\ref{fig:fig4}(e). Despite the degradation in fidelity with deviation of $J_{test}$ from $J_{train} = t$, our model is still able to predict the desired $\mathbf{q}$ vector with fidelity above 0.8 for $J_{train}-J_{test}\ge 0.6t$. However, when the NN is trained with the smaller value of $J_{train}=0.2t$, the fidelity drops down to $\mathcal{F}_{|\mathbf{q}|}=0.86$ and $\mathcal F_{\theta}=0.64$ already for a deviation $J_{train}-J_{test} = -0.2t$. A weaker exchange coupling diminishes the impact of the proximity effect and, as a result, reduces the conductance signatures it produces. Thus, a sufficiently strong exchange coupling, in particular stronger than the noise, is crucial for implementing our strategy experimentally. In realistic experimental scenarios, we expect the exchange coupling to be around $J=0.05 - 1.0$~meV \cite{tang2020, campbell2022, regan2024, kiese2022}.

Finally, it is worth noting that our methodology relies on several assumptions. We calculate conductance within a single-particle framework, omitting strongly interacting many-body effects, which are beyond the scope of this study. To resolve small variations in conductance induced by SSMs, we assume cryogenic temperatures below the critical temperature of the spin spiral, which are accessible with modern cryostats \cite{lee2018, konig2007}. We also take the exchange coupling $J$ to be constant and the spin spiral to be perfectly ordered,
assuming that a single magnetic domain is present. 
Furthermore, our methodology assumes that the spin spiral remains unaffected by the magnetic field, which requires spin-spiral internal exchange couplings
stronger than the Zeeman effect. Finally, we work in the ballistic phase-coherent transport regime, which is realistically achievable in high-quality nanoribbons with lengths below 100~nm \cite{mukhopadhyay2021, schmidt2014, talkington2025, Guerrero2025}. 

\section{Conclusions}
\label{sec:conclusions}
We have demonstrated a ML methodology that leverages electronic transport data to extract the $\mathbf{q}$ vector of a SSM. 
In particular, our strategy exploits the interplay between electronic transport,
magnetic field, electronic density, and exchange proximity
to directly infer the $\mathbf{q}$ vector of a SSM.
From a physical perspective, our algorithm
extracts the subtle impact of the spiral exchange coupling in Hofstadter
transport spectra, using electronic transport as a probe of noncollinear
magnetism.
We also demonstrated that our methodology is substantially resilient to noise in electronic transport, a crucial condition for experimental feasibility.
These results provide an experimentally accessible strategy for Hamiltonian learning of 2D SSMs in moir\'e superlattices of twisted vdW materials,
establishing electronic transport as a flexible tool for identifying
nontrivial spin textures.

\FloatBarrier

\begin{acknowledgments}
 This work was supported by the Research Council of  Finland Flagship Programme (PREIN, Quantum), ERC (Grant No. 834742), ERC-2024-CoG ULTRATWISTROICS (Grant No.~101170477) the Research Council of Finland (360411, 359009, 365686, 367808, and 374168), the School of Electrical Engineering (Aalto University), InstituteQ, the Finnish Quantum Flagship, the Nokia Foundation, and the Finnish Centre of Excellence in Quantum Materials QMAT. We acknowledge the computational resources provided by the Aalto Science-IT project.
\end{acknowledgments}

\appendix*
\section{Principal components analysis and training process}
\label{appendixA}
The dimensionality of the initial dataset has been reduced by applying PCA. The number of PCs is chosen based on the cumulative variance and the maximization of the final fidelity. Figure~\ref{fig:fig6}(a) shows the cumulative variance as a function of the number of PCs.  Thus, for the prediction results in Fig.~\ref{fig:fig3}, we used 500 PCs given by the cumulative variance of 0.99 instead of the 1300 initial components.

In this paper, we employed a supervised ML algorithm implemented with a Keras/TensorFlow feedforward NN with two fully connected hidden layers, each containing 100 neurons. The NN was trained using Adam optimizer with a learning rate of $10^{-3}$ and ReLU activation function. The training process was performed with a batch size of 16 in 100 epochs by optimizing the mean squared error (MSE) loss for training and validation datasets. Figure~\ref{fig:fig6}(b) shows the history of the training process. The consistent decrease in both training and validation MSE indicates no evidence of overfitting.

\bibliography{apssamp}

\begin{thebibliography}{56}%
\makeatletter
\providecommand \@ifxundefined [1]{%
 \@ifx{#1\undefined}
}%
\providecommand \@ifnum [1]{%
 \ifnum #1\expandafter \@firstoftwo
 \else \expandafter \@secondoftwo
 \fi
}%
\providecommand \@ifx [1]{%
 \ifx #1\expandafter \@firstoftwo
 \else \expandafter \@secondoftwo
 \fi
}%
\providecommand \natexlab [1]{#1}%
\providecommand \enquote  [1]{``#1''}%
\providecommand \bibnamefont  [1]{#1}%
\providecommand \bibfnamefont [1]{#1}%
\providecommand \citenamefont [1]{#1}%
\providecommand \href@noop [0]{\@secondoftwo}%
\providecommand \href [0]{\begingroup \@sanitize@url \@href}%
\providecommand \@href[1]{\@@startlink{#1}\@@href}%
\providecommand \@@href[1]{\endgroup#1\@@endlink}%
\providecommand \@sanitize@url [0]{\catcode `\\12\catcode `\$12\catcode `\&12\catcode `\#12\catcode `\^12\catcode `\_12\catcode `\%12\relax}%
\providecommand \@@startlink[1]{}%
\providecommand \@@endlink[0]{}%
\providecommand \url  [0]{\begingroup\@sanitize@url \@url }%
\providecommand \@url [1]{\endgroup\@href {#1}{\urlprefix }}%
\providecommand \urlprefix  [0]{URL }%
\providecommand \Eprint [0]{\href }%
\providecommand \doibase [0]{https://doi.org/}%
\providecommand \selectlanguage [0]{\@gobble}%
\providecommand \bibinfo  [0]{\@secondoftwo}%
\providecommand \bibfield  [0]{\@secondoftwo}%
\providecommand \translation [1]{[#1]}%
\providecommand \BibitemOpen [0]{}%
\providecommand \bibitemStop [0]{}%
\providecommand \bibitemNoStop [0]{.\EOS\space}%
\providecommand \EOS [0]{\spacefactor3000\relax}%
\providecommand \BibitemShut  [1]{\csname bibitem#1\endcsname}%
\let\auto@bib@innerbib\@empty
\bibitem [{\citenamefont {Mak}\ \emph {et~al.}(2019)\citenamefont {Mak}, \citenamefont {Shan},\ and\ \citenamefont {Ralph}}]{mak2019}%
  \BibitemOpen
  \bibfield  {author} {\bibinfo {author} {\bibfnamefont {K.~F.}\ \bibnamefont {Mak}}, \bibinfo {author} {\bibfnamefont {J.}~\bibnamefont {Shan}},\ and\ \bibinfo {author} {\bibfnamefont {D.~C.}\ \bibnamefont {Ralph}},\ }\bibfield  {title} {\bibinfo {title} {Probing and controlling magnetic states in 2{D} layered magnetic materials},\ }\href {https://doi.org/10.1038/s42254-019-0110-y} {\bibfield  {journal} {\bibinfo  {journal} {Nat. Rev. Phys.}\ }\textbf {\bibinfo {volume} {1}},\ \bibinfo {pages} {646} (\bibinfo {year} {2019})}\BibitemShut {NoStop}%
\bibitem [{\citenamefont {Song}\ \emph {et~al.}(2025)\citenamefont {Song}, \citenamefont {Stavrić}, \citenamefont {Barone}, \citenamefont {Droghetti}, \citenamefont {Antonenko}, \citenamefont {Venderbos}, \citenamefont {Occhialini}, \citenamefont {Ilyas}, \citenamefont {Erge\c{c}en}, \citenamefont {Gedik}, \citenamefont {Cheong}, \citenamefont {Fernandes}, \citenamefont {Picozzi},\ and\ \citenamefont {Comin}}]{Song2025}%
  \BibitemOpen
  \bibfield  {author} {\bibinfo {author} {\bibfnamefont {Q.}~\bibnamefont {Song}}, \bibinfo {author} {\bibfnamefont {S.}~\bibnamefont {Stavrić}}, \bibinfo {author} {\bibfnamefont {P.}~\bibnamefont {Barone}}, \bibinfo {author} {\bibfnamefont {A.}~\bibnamefont {Droghetti}}, \bibinfo {author} {\bibfnamefont {D.~S.}\ \bibnamefont {Antonenko}}, \bibinfo {author} {\bibfnamefont {J.~W.~F.}\ \bibnamefont {Venderbos}}, \bibinfo {author} {\bibfnamefont {C.~A.}\ \bibnamefont {Occhialini}}, \bibinfo {author} {\bibfnamefont {B.}~\bibnamefont {Ilyas}}, \bibinfo {author} {\bibfnamefont {E.}~\bibnamefont {Erge\c{c}en}}, \bibinfo {author} {\bibfnamefont {N.}~\bibnamefont {Gedik}}, \bibinfo {author} {\bibfnamefont {S.-W.}\ \bibnamefont {Cheong}}, \bibinfo {author} {\bibfnamefont {R.~M.}\ \bibnamefont {Fernandes}}, \bibinfo {author} {\bibfnamefont {S.}~\bibnamefont {Picozzi}},\ and\ \bibinfo {author} {\bibfnamefont {R.}~\bibnamefont {Comin}},\ }\bibfield  {title} {\bibinfo {title} {Electrical switching of a p-wave magnet},\
  }\href {https://doi.org/10.1038/s41586-025-09034-7} {\bibfield  {journal} {\bibinfo  {journal} {Nature}\ }\textbf {\bibinfo {volume} {642}},\ \bibinfo {pages} {64–70} (\bibinfo {year} {2025})}\BibitemShut {NoStop}%
\bibitem [{\citenamefont {Wang}\ \emph {et~al.}(2024)\citenamefont {Wang}, \citenamefont {Zhao}, \citenamefont {Yao}, \citenamefont {Liu}, \citenamefont {Cheng}, \citenamefont {Zhang}, \citenamefont {Feng}, \citenamefont {Ma}, \citenamefont {Zhao}, \citenamefont {Sun}, \citenamefont {Wu},\ and\ \citenamefont {Chen}}]{Wang2024}%
  \BibitemOpen
  \bibfield  {author} {\bibinfo {author} {\bibfnamefont {Y.}~\bibnamefont {Wang}}, \bibinfo {author} {\bibfnamefont {X.}~\bibnamefont {Zhao}}, \bibinfo {author} {\bibfnamefont {L.}~\bibnamefont {Yao}}, \bibinfo {author} {\bibfnamefont {H.}~\bibnamefont {Liu}}, \bibinfo {author} {\bibfnamefont {P.}~\bibnamefont {Cheng}}, \bibinfo {author} {\bibfnamefont {Y.}~\bibnamefont {Zhang}}, \bibinfo {author} {\bibfnamefont {B.}~\bibnamefont {Feng}}, \bibinfo {author} {\bibfnamefont {F.}~\bibnamefont {Ma}}, \bibinfo {author} {\bibfnamefont {J.}~\bibnamefont {Zhao}}, \bibinfo {author} {\bibfnamefont {J.}~\bibnamefont {Sun}}, \bibinfo {author} {\bibfnamefont {K.}~\bibnamefont {Wu}},\ and\ \bibinfo {author} {\bibfnamefont {L.}~\bibnamefont {Chen}},\ }\bibfield  {title} {\bibinfo {title} {Orientation-selective spin-polarized edge states in monolayer {N}i{I}$_2$},\ }\href {https://doi.org/10.1038/s41467-024-55372-x} {\bibfield  {journal} {\bibinfo  {journal} {Nat. Commun.}\ }\textbf {\bibinfo {volume} {15}},\ \bibinfo {pages}
  {10916} (\bibinfo {year} {2024})}\BibitemShut {NoStop}%
\bibitem [{\citenamefont {Miao}\ \emph {et~al.}(2025)\citenamefont {Miao}, \citenamefont {Liu}, \citenamefont {Zhang}, \citenamefont {Zhou}, \citenamefont {Wang}, \citenamefont {Wang}, \citenamefont {Ji},\ and\ \citenamefont {Fu}}]{Miao2025}%
  \BibitemOpen
  \bibfield  {author} {\bibinfo {author} {\bibfnamefont {M.-P.}\ \bibnamefont {Miao}}, \bibinfo {author} {\bibfnamefont {N.}~\bibnamefont {Liu}}, \bibinfo {author} {\bibfnamefont {W.-H.}\ \bibnamefont {Zhang}}, \bibinfo {author} {\bibfnamefont {J.-W.}\ \bibnamefont {Zhou}}, \bibinfo {author} {\bibfnamefont {D.-B.}\ \bibnamefont {Wang}}, \bibinfo {author} {\bibfnamefont {C.}~\bibnamefont {Wang}}, \bibinfo {author} {\bibfnamefont {W.}~\bibnamefont {Ji}},\ and\ \bibinfo {author} {\bibfnamefont {Y.-S.}\ \bibnamefont {Fu}},\ }\bibfield  {title} {\bibinfo {title} {Spin-resolved imaging of atomic-scale helimagnetism in mono- and bilayer {N}i{I}$_2$},\ }\href {https://doi.org/10.1073/pnas.2422868122} {\bibfield  {journal} {\bibinfo  {journal} {Proc. Natl. Acad. Sci. U.S.A.}\ }\textbf {\bibinfo {volume} {122}},\ \bibinfo {pages} {e2422868122} (\bibinfo {year} {2025})}\BibitemShut {NoStop}%
\bibitem [{\citenamefont {Nigmatulin}\ \emph {et~al.}(2025)\citenamefont {Nigmatulin}, \citenamefont {Lado},\ and\ \citenamefont {Sun}}]{Nigmatulin2025}%
  \BibitemOpen
  \bibfield  {author} {\bibinfo {author} {\bibfnamefont {F.}~\bibnamefont {Nigmatulin}}, \bibinfo {author} {\bibfnamefont {J.~L.}\ \bibnamefont {Lado}},\ and\ \bibinfo {author} {\bibfnamefont {Z.}~\bibnamefont {Sun}},\ }\bibfield  {title} {\bibinfo {title} {Electrical probe of spin-spiral order in quantum spin {H}all/spin-spiral magnet van der {W}aals heterostructures},\ }\href {https://doi.org/10.1103/2v3v-3l7p} {\bibfield  {journal} {\bibinfo  {journal} {Phys. Rev. B}\ }\textbf {\bibinfo {volume} {112}},\ \bibinfo {pages} {024430} (\bibinfo {year} {2025})}\BibitemShut {NoStop}%
\bibitem [{\citenamefont {Amini}\ \emph {et~al.}(2024)\citenamefont {Amini}, \citenamefont {Fumega}, \citenamefont {González-Herrero}, \citenamefont {Vaňo}, \citenamefont {Kezilebieke}, \citenamefont {Lado},\ and\ \citenamefont {Liljeroth}}]{Amini2024}%
  \BibitemOpen
  \bibfield  {author} {\bibinfo {author} {\bibfnamefont {M.}~\bibnamefont {Amini}}, \bibinfo {author} {\bibfnamefont {A.~O.}\ \bibnamefont {Fumega}}, \bibinfo {author} {\bibfnamefont {H.}~\bibnamefont {González-Herrero}}, \bibinfo {author} {\bibfnamefont {V.}~\bibnamefont {Vaňo}}, \bibinfo {author} {\bibfnamefont {S.}~\bibnamefont {Kezilebieke}}, \bibinfo {author} {\bibfnamefont {J.~L.}\ \bibnamefont {Lado}},\ and\ \bibinfo {author} {\bibfnamefont {P.}~\bibnamefont {Liljeroth}},\ }\bibfield  {title} {\bibinfo {title} {Atomic-scale visualization of multiferroicity in monolayer {N}i{I}$_2$},\ }\href {https://doi.org/https://doi.org/10.1002/adma.202311342} {\bibfield  {journal} {\bibinfo  {journal} {Adv. Mater.}\ }\textbf {\bibinfo {volume} {36}},\ \bibinfo {pages} {2311342} (\bibinfo {year} {2024})}\BibitemShut {NoStop}%
\bibitem [{\citenamefont {Wang}\ \emph {et~al.}(2017)\citenamefont {Wang}, \citenamefont {Paesani}, \citenamefont {Santagati}, \citenamefont {Knauer}, \citenamefont {Gentile}, \citenamefont {Wiebe}, \citenamefont {Petruzzella}, \citenamefont {O'Brien}, \citenamefont {Rarity}, \citenamefont {Laing},\ and\ \citenamefont {Thompson}}]{Wang2017}%
  \BibitemOpen
  \bibfield  {author} {\bibinfo {author} {\bibfnamefont {J.}~\bibnamefont {Wang}}, \bibinfo {author} {\bibfnamefont {S.}~\bibnamefont {Paesani}}, \bibinfo {author} {\bibfnamefont {R.}~\bibnamefont {Santagati}}, \bibinfo {author} {\bibfnamefont {S.}~\bibnamefont {Knauer}}, \bibinfo {author} {\bibfnamefont {A.~A.}\ \bibnamefont {Gentile}}, \bibinfo {author} {\bibfnamefont {N.}~\bibnamefont {Wiebe}}, \bibinfo {author} {\bibfnamefont {M.}~\bibnamefont {Petruzzella}}, \bibinfo {author} {\bibfnamefont {J.~L.}\ \bibnamefont {O'Brien}}, \bibinfo {author} {\bibfnamefont {J.~G.}\ \bibnamefont {Rarity}}, \bibinfo {author} {\bibfnamefont {A.}~\bibnamefont {Laing}},\ and\ \bibinfo {author} {\bibfnamefont {M.~G.}\ \bibnamefont {Thompson}},\ }\bibfield  {title} {\bibinfo {title} {Experimental quantum {H}amiltonian learning},\ }\href {https://doi.org/10.1038/nphys4074} {\bibfield  {journal} {\bibinfo  {journal} {Nat. Phys.}\ }\textbf {\bibinfo {volume} {13}},\ \bibinfo {pages} {551} (\bibinfo {year} {2017})}\BibitemShut
  {NoStop}%
\bibitem [{\citenamefont {Valenti}\ \emph {et~al.}(2022)\citenamefont {Valenti}, \citenamefont {Jin}, \citenamefont {L\'eonard}, \citenamefont {Huber},\ and\ \citenamefont {Greplova}}]{Valenti2022}%
  \BibitemOpen
  \bibfield  {author} {\bibinfo {author} {\bibfnamefont {A.}~\bibnamefont {Valenti}}, \bibinfo {author} {\bibfnamefont {G.}~\bibnamefont {Jin}}, \bibinfo {author} {\bibfnamefont {J.}~\bibnamefont {L\'eonard}}, \bibinfo {author} {\bibfnamefont {S.~D.}\ \bibnamefont {Huber}},\ and\ \bibinfo {author} {\bibfnamefont {E.}~\bibnamefont {Greplova}},\ }\bibfield  {title} {\bibinfo {title} {Scalable {H}amiltonian learning for large-scale out-of-equilibrium quantum dynamics},\ }\href {https://doi.org/10.1103/PhysRevA.105.023302} {\bibfield  {journal} {\bibinfo  {journal} {Phys. Rev. A}\ }\textbf {\bibinfo {volume} {105}},\ \bibinfo {pages} {023302} (\bibinfo {year} {2022})}\BibitemShut {NoStop}%
\bibitem [{\citenamefont {Koch}\ \emph {et~al.}(2025)\citenamefont {Koch}, \citenamefont {Drost}, \citenamefont {Liljeroth},\ and\ \citenamefont {Lado}}]{Koch2025}%
  \BibitemOpen
  \bibfield  {author} {\bibinfo {author} {\bibfnamefont {R.}~\bibnamefont {Koch}}, \bibinfo {author} {\bibfnamefont {R.}~\bibnamefont {Drost}}, \bibinfo {author} {\bibfnamefont {P.}~\bibnamefont {Liljeroth}},\ and\ \bibinfo {author} {\bibfnamefont {J.~L.}\ \bibnamefont {Lado}},\ }\bibfield  {title} {\bibinfo {title} {Hamiltonian learning of triplon excitations in an artificial nanoscale molecular quantum magnet},\ }\href {https://doi.org/10.1021/acs.nanolett.5c02502} {\bibfield  {journal} {\bibinfo  {journal} {Nano Lett.}\ }\textbf {\bibinfo {volume} {25}},\ \bibinfo {pages} {13435} (\bibinfo {year} {2025})}\BibitemShut {NoStop}%
\bibitem [{\citenamefont {Bairey}\ \emph {et~al.}(2019)\citenamefont {Bairey}, \citenamefont {Arad},\ and\ \citenamefont {Lindner}}]{Barey2019}%
  \BibitemOpen
  \bibfield  {author} {\bibinfo {author} {\bibfnamefont {E.}~\bibnamefont {Bairey}}, \bibinfo {author} {\bibfnamefont {I.}~\bibnamefont {Arad}},\ and\ \bibinfo {author} {\bibfnamefont {N.~H.}\ \bibnamefont {Lindner}},\ }\bibfield  {title} {\bibinfo {title} {Learning a local {H}amiltonian from local measurements},\ }\href {https://doi.org/10.1103/PhysRevLett.122.020504} {\bibfield  {journal} {\bibinfo  {journal} {Phys. Rev. Lett.}\ }\textbf {\bibinfo {volume} {122}},\ \bibinfo {pages} {020504} (\bibinfo {year} {2019})}\BibitemShut {NoStop}%
\bibitem [{\citenamefont {Gebhart}\ \emph {et~al.}(2023)\citenamefont {Gebhart}, \citenamefont {Santagati}, \citenamefont {Gentile}, \citenamefont {Gauger}, \citenamefont {Craig}, \citenamefont {Ares}, \citenamefont {Banchi}, \citenamefont {Marquardt}, \citenamefont {Pezz{\`e}},\ and\ \citenamefont {Bonato}}]{Gebhart2023}%
  \BibitemOpen
  \bibfield  {author} {\bibinfo {author} {\bibfnamefont {V.}~\bibnamefont {Gebhart}}, \bibinfo {author} {\bibfnamefont {R.}~\bibnamefont {Santagati}}, \bibinfo {author} {\bibfnamefont {A.~A.}\ \bibnamefont {Gentile}}, \bibinfo {author} {\bibfnamefont {E.~M.}\ \bibnamefont {Gauger}}, \bibinfo {author} {\bibfnamefont {D.}~\bibnamefont {Craig}}, \bibinfo {author} {\bibfnamefont {N.}~\bibnamefont {Ares}}, \bibinfo {author} {\bibfnamefont {L.}~\bibnamefont {Banchi}}, \bibinfo {author} {\bibfnamefont {F.}~\bibnamefont {Marquardt}}, \bibinfo {author} {\bibfnamefont {L.}~\bibnamefont {Pezz{\`e}}},\ and\ \bibinfo {author} {\bibfnamefont {C.}~\bibnamefont {Bonato}},\ }\bibfield  {title} {\bibinfo {title} {Learning quantum systems},\ }\href {https://doi.org/10.1038/s42254-022-00552-1} {\bibfield  {journal} {\bibinfo  {journal} {Nat. Rev. Phys.}\ }\textbf {\bibinfo {volume} {5}},\ \bibinfo {pages} {141} (\bibinfo {year} {2023})}\BibitemShut {NoStop}%
\bibitem [{\citenamefont {Lupi}\ and\ \citenamefont {Lado}(2025)}]{Lupi2025}%
  \BibitemOpen
  \bibfield  {author} {\bibinfo {author} {\bibfnamefont {G.}~\bibnamefont {Lupi}}\ and\ \bibinfo {author} {\bibfnamefont {J.~L.}\ \bibnamefont {Lado}},\ }\bibfield  {title} {\bibinfo {title} {Hamiltonian-learning quantum magnets with nonlocal impurity tomography},\ }\href {https://doi.org/10.1103/PhysRevApplied.23.054077} {\bibfield  {journal} {\bibinfo  {journal} {Phys. Rev. Appl.}\ }\textbf {\bibinfo {volume} {23}},\ \bibinfo {pages} {054077} (\bibinfo {year} {2025})}\BibitemShut {NoStop}%
\bibitem [{\citenamefont {van Driel}\ \emph {et~al.}()\citenamefont {van Driel}, \citenamefont {Koch}, \citenamefont {Sietses}, \citenamefont {ten Haaf}, \citenamefont {Liu}, \citenamefont {Zatelli}, \citenamefont {Roovers}, \citenamefont {Bordin}, \citenamefont {van Loo}, \citenamefont {Wang}, \citenamefont {Wolff}, \citenamefont {Mazur}, \citenamefont {Dvir}, \citenamefont {Kulesh}, \citenamefont {Wang}, \citenamefont {Bozkurt}, \citenamefont {Gazibegovic}, \citenamefont {Badawy}, \citenamefont {Bakkers}, \citenamefont {Wimmer}, \citenamefont {Goswami}, \citenamefont {Lado}, \citenamefont {Kouwenhoven},\ and\ \citenamefont {Greplova}}]{vandriel2024}%
  \BibitemOpen
  \bibfield  {author} {\bibinfo {author} {\bibfnamefont {D.}~\bibnamefont {van Driel}}, \bibinfo {author} {\bibfnamefont {R.}~\bibnamefont {Koch}}, \bibinfo {author} {\bibfnamefont {V.~P.~M.}\ \bibnamefont {Sietses}}, \bibinfo {author} {\bibfnamefont {S.~L.~D.}\ \bibnamefont {ten Haaf}}, \bibinfo {author} {\bibfnamefont {C.-X.}\ \bibnamefont {Liu}}, \bibinfo {author} {\bibfnamefont {F.}~\bibnamefont {Zatelli}}, \bibinfo {author} {\bibfnamefont {B.}~\bibnamefont {Roovers}}, \bibinfo {author} {\bibfnamefont {A.}~\bibnamefont {Bordin}}, \bibinfo {author} {\bibfnamefont {N.}~\bibnamefont {van Loo}}, \bibinfo {author} {\bibfnamefont {G.}~\bibnamefont {Wang}}, \bibinfo {author} {\bibfnamefont {J.~C.}\ \bibnamefont {Wolff}}, \bibinfo {author} {\bibfnamefont {G.~P.}\ \bibnamefont {Mazur}}, \bibinfo {author} {\bibfnamefont {T.}~\bibnamefont {Dvir}}, \bibinfo {author} {\bibfnamefont {I.}~\bibnamefont {Kulesh}}, \bibinfo {author} {\bibfnamefont {Q.}~\bibnamefont {Wang}}, \bibinfo {author} {\bibfnamefont {A.~M.}\
  \bibnamefont {Bozkurt}}, \bibinfo {author} {\bibfnamefont {S.}~\bibnamefont {Gazibegovic}}, \bibinfo {author} {\bibfnamefont {G.}~\bibnamefont {Badawy}}, \bibinfo {author} {\bibfnamefont {E.~P. A.~M.}\ \bibnamefont {Bakkers}}, \bibinfo {author} {\bibfnamefont {M.}~\bibnamefont {Wimmer}}, \bibinfo {author} {\bibfnamefont {S.}~\bibnamefont {Goswami}}, \bibinfo {author} {\bibfnamefont {J.~L.}\ \bibnamefont {Lado}}, \bibinfo {author} {\bibfnamefont {L.~P.}\ \bibnamefont {Kouwenhoven}},\ and\ \bibinfo {author} {\bibfnamefont {E.}~\bibnamefont {Greplova}},\ }\bibfield  {title} {\bibinfo {title} {Cross-platform autonomous control of minimal {K}itaev chains},\ }\href@noop {} {\ }\Eprint {https://arxiv.org/abs/2405.04596} {arXiv:2405.04596} \BibitemShut {NoStop}%
\bibitem [{\citenamefont {Karjalainen}\ \emph {et~al.}(2023)\citenamefont {Karjalainen}, \citenamefont {Lippo}, \citenamefont {Chen}, \citenamefont {Koch}, \citenamefont {Fumega},\ and\ \citenamefont {Lado}}]{Karjalainen2023}%
  \BibitemOpen
  \bibfield  {author} {\bibinfo {author} {\bibfnamefont {N.}~\bibnamefont {Karjalainen}}, \bibinfo {author} {\bibfnamefont {Z.}~\bibnamefont {Lippo}}, \bibinfo {author} {\bibfnamefont {G.}~\bibnamefont {Chen}}, \bibinfo {author} {\bibfnamefont {R.}~\bibnamefont {Koch}}, \bibinfo {author} {\bibfnamefont {A.~O.}\ \bibnamefont {Fumega}},\ and\ \bibinfo {author} {\bibfnamefont {J.~L.}\ \bibnamefont {Lado}},\ }\bibfield  {title} {\bibinfo {title} {Hamiltonian inference from dynamical excitations in confined quantum magnets},\ }\href {https://doi.org/10.1103/PhysRevApplied.20.024054} {\bibfield  {journal} {\bibinfo  {journal} {Phys. Rev. Appl.}\ }\textbf {\bibinfo {volume} {20}},\ \bibinfo {pages} {024054} (\bibinfo {year} {2023})}\BibitemShut {NoStop}%
\bibitem [{\citenamefont {Karjalainen}\ \emph {et~al.}()\citenamefont {Karjalainen}, \citenamefont {Lupi}, \citenamefont {Koch}, \citenamefont {Fumega},\ and\ \citenamefont {Lado}}]{Karjalainen2025}%
  \BibitemOpen
  \bibfield  {author} {\bibinfo {author} {\bibfnamefont {N.}~\bibnamefont {Karjalainen}}, \bibinfo {author} {\bibfnamefont {G.}~\bibnamefont {Lupi}}, \bibinfo {author} {\bibfnamefont {R.}~\bibnamefont {Koch}}, \bibinfo {author} {\bibfnamefont {A.~O.}\ \bibnamefont {Fumega}},\ and\ \bibinfo {author} {\bibfnamefont {J.~L.}\ \bibnamefont {Lado}},\ }\bibfield  {title} {\bibinfo {title} {Hamiltonian learning quantum magnets with dynamical impurity tomography},\ }\href@noop {} {\ }\Eprint {https://arxiv.org/abs/2510.18613} {arXiv:2510.18613} \BibitemShut {NoStop}%
\bibitem [{\citenamefont {Koch}\ and\ \citenamefont {Lado}(2022)}]{Koch2022}%
  \BibitemOpen
  \bibfield  {author} {\bibinfo {author} {\bibfnamefont {R.}~\bibnamefont {Koch}}\ and\ \bibinfo {author} {\bibfnamefont {J.~L.}\ \bibnamefont {Lado}},\ }\bibfield  {title} {\bibinfo {title} {Designing quantum many-body matter with conditional generative adversarial networks},\ }\href {https://doi.org/10.1103/PhysRevResearch.4.033223} {\bibfield  {journal} {\bibinfo  {journal} {Phys. Rev. Res.}\ }\textbf {\bibinfo {volume} {4}},\ \bibinfo {pages} {033223} (\bibinfo {year} {2022})}\BibitemShut {NoStop}%
\bibitem [{\citenamefont {Koch}\ \emph {et~al.}(2023)\citenamefont {Koch}, \citenamefont {van Driel}, \citenamefont {Bordin}, \citenamefont {Lado},\ and\ \citenamefont {Greplova}}]{Koch2023}%
  \BibitemOpen
  \bibfield  {author} {\bibinfo {author} {\bibfnamefont {R.}~\bibnamefont {Koch}}, \bibinfo {author} {\bibfnamefont {D.}~\bibnamefont {van Driel}}, \bibinfo {author} {\bibfnamefont {A.}~\bibnamefont {Bordin}}, \bibinfo {author} {\bibfnamefont {J.~L.}\ \bibnamefont {Lado}},\ and\ \bibinfo {author} {\bibfnamefont {E.}~\bibnamefont {Greplova}},\ }\bibfield  {title} {\bibinfo {title} {Adversarial {H}amiltonian learning of quantum dots in a minimal {K}itaev chain},\ }\href {https://doi.org/10.1103/PhysRevApplied.20.044081} {\bibfield  {journal} {\bibinfo  {journal} {Phys. Rev. Appl.}\ }\textbf {\bibinfo {volume} {20}},\ \bibinfo {pages} {044081} (\bibinfo {year} {2023})}\BibitemShut {NoStop}%
\bibitem [{\citenamefont {Che}\ \emph {et~al.}(2021)\citenamefont {Che}, \citenamefont {Wei}, \citenamefont {Huang}, \citenamefont {Zhao}, \citenamefont {Xue}, \citenamefont {Nie}, \citenamefont {Li}, \citenamefont {Lu},\ and\ \citenamefont {Xin}}]{Che2021}%
  \BibitemOpen
  \bibfield  {author} {\bibinfo {author} {\bibfnamefont {L.}~\bibnamefont {Che}}, \bibinfo {author} {\bibfnamefont {C.}~\bibnamefont {Wei}}, \bibinfo {author} {\bibfnamefont {Y.}~\bibnamefont {Huang}}, \bibinfo {author} {\bibfnamefont {D.}~\bibnamefont {Zhao}}, \bibinfo {author} {\bibfnamefont {S.}~\bibnamefont {Xue}}, \bibinfo {author} {\bibfnamefont {X.}~\bibnamefont {Nie}}, \bibinfo {author} {\bibfnamefont {J.}~\bibnamefont {Li}}, \bibinfo {author} {\bibfnamefont {D.}~\bibnamefont {Lu}},\ and\ \bibinfo {author} {\bibfnamefont {T.}~\bibnamefont {Xin}},\ }\bibfield  {title} {\bibinfo {title} {Learning quantum {H}amiltonians from single-qubit measurements},\ }\href {https://doi.org/10.1103/PhysRevResearch.3.023246} {\bibfield  {journal} {\bibinfo  {journal} {Phys. Rev. Res.}\ }\textbf {\bibinfo {volume} {3}},\ \bibinfo {pages} {023246} (\bibinfo {year} {2021})}\BibitemShut {NoStop}%
\bibitem [{\citenamefont {Simard}\ \emph {et~al.}(2025)\citenamefont {Simard}, \citenamefont {Dawid}, \citenamefont {Tindall}, \citenamefont {Ferrero}, \citenamefont {Sengupta},\ and\ \citenamefont {Georges}}]{Simard2025}%
  \BibitemOpen
  \bibfield  {author} {\bibinfo {author} {\bibfnamefont {O.}~\bibnamefont {Simard}}, \bibinfo {author} {\bibfnamefont {A.}~\bibnamefont {Dawid}}, \bibinfo {author} {\bibfnamefont {J.}~\bibnamefont {Tindall}}, \bibinfo {author} {\bibfnamefont {M.}~\bibnamefont {Ferrero}}, \bibinfo {author} {\bibfnamefont {A.~M.}\ \bibnamefont {Sengupta}},\ and\ \bibinfo {author} {\bibfnamefont {A.}~\bibnamefont {Georges}},\ }\bibfield  {title} {\bibinfo {title} {Learning interactions between {R}ydberg atoms},\ }\href {https://doi.org/10.1103/f58h-zxs3} {\bibfield  {journal} {\bibinfo  {journal} {PRX Quantum}\ }\textbf {\bibinfo {volume} {6}},\ \bibinfo {pages} {030324} (\bibinfo {year} {2025})}\BibitemShut {NoStop}%
\bibitem [{\citenamefont {Wang}\ \emph {et~al.}(2020)\citenamefont {Wang}, \citenamefont {Wei}, \citenamefont {Yuan}, \citenamefont {Tian}, \citenamefont {Cao}, \citenamefont {Zhao}, \citenamefont {Zhang}, \citenamefont {Zhou}, \citenamefont {Song}, \citenamefont {Xue},\ and\ \citenamefont {Yang}}]{Wang2020}%
  \BibitemOpen
  \bibfield  {author} {\bibinfo {author} {\bibfnamefont {D.}~\bibnamefont {Wang}}, \bibinfo {author} {\bibfnamefont {S.}~\bibnamefont {Wei}}, \bibinfo {author} {\bibfnamefont {A.}~\bibnamefont {Yuan}}, \bibinfo {author} {\bibfnamefont {F.}~\bibnamefont {Tian}}, \bibinfo {author} {\bibfnamefont {K.}~\bibnamefont {Cao}}, \bibinfo {author} {\bibfnamefont {Q.}~\bibnamefont {Zhao}}, \bibinfo {author} {\bibfnamefont {Y.}~\bibnamefont {Zhang}}, \bibinfo {author} {\bibfnamefont {C.}~\bibnamefont {Zhou}}, \bibinfo {author} {\bibfnamefont {X.}~\bibnamefont {Song}}, \bibinfo {author} {\bibfnamefont {D.}~\bibnamefont {Xue}},\ and\ \bibinfo {author} {\bibfnamefont {S.}~\bibnamefont {Yang}},\ }\bibfield  {title} {\bibinfo {title} {Machine learning magnetic parameters from spin configurations},\ }\href {https://doi.org/https://doi.org/10.1002/advs.202000566} {\bibfield  {journal} {\bibinfo  {journal} {Adv. Sci.}\ }\textbf {\bibinfo {volume} {7}},\ \bibinfo {pages} {2000566} (\bibinfo {year} {2020})}\BibitemShut {NoStop}%
\bibitem [{\citenamefont {Salcedo-Gallo}\ \emph {et~al.}(2020)\citenamefont {Salcedo-Gallo}, \citenamefont {Galindo-González},\ and\ \citenamefont {Restrepo-Parra}}]{SALCEDOGALLO2020}%
  \BibitemOpen
  \bibfield  {author} {\bibinfo {author} {\bibfnamefont {J.}~\bibnamefont {Salcedo-Gallo}}, \bibinfo {author} {\bibfnamefont {C.}~\bibnamefont {Galindo-González}},\ and\ \bibinfo {author} {\bibfnamefont {E.}~\bibnamefont {Restrepo-Parra}},\ }\bibfield  {title} {\bibinfo {title} {Deep learning approach for image classification of magnetic phases in chiral magnets},\ }\href {https://doi.org/https://doi.org/10.1016/j.jmmm.2020.166482} {\bibfield  {journal} {\bibinfo  {journal} {J. Magn. Magn. Mater.}\ }\textbf {\bibinfo {volume} {501}},\ \bibinfo {pages} {166482} (\bibinfo {year} {2020})}\BibitemShut {NoStop}%
\bibitem [{\citenamefont {Iakovlev}\ \emph {et~al.}(2018)\citenamefont {Iakovlev}, \citenamefont {Sotnikov},\ and\ \citenamefont {Mazurenko}}]{Iakovlev2018}%
  \BibitemOpen
  \bibfield  {author} {\bibinfo {author} {\bibfnamefont {I.~A.}\ \bibnamefont {Iakovlev}}, \bibinfo {author} {\bibfnamefont {O.~M.}\ \bibnamefont {Sotnikov}},\ and\ \bibinfo {author} {\bibfnamefont {V.~V.}\ \bibnamefont {Mazurenko}},\ }\bibfield  {title} {\bibinfo {title} {Supervised learning approach for recognizing magnetic skyrmion phases},\ }\href {https://doi.org/10.1103/PhysRevB.98.174411} {\bibfield  {journal} {\bibinfo  {journal} {Phys. Rev. B}\ }\textbf {\bibinfo {volume} {98}},\ \bibinfo {pages} {174411} (\bibinfo {year} {2018})}\BibitemShut {NoStop}%
\bibitem [{\citenamefont {Feng}\ \emph {et~al.}(2024)\citenamefont {Feng}, \citenamefont {Guan}, \citenamefont {Wu}, \citenamefont {Wu},\ and\ \citenamefont {Song}}]{Feng2024}%
  \BibitemOpen
  \bibfield  {author} {\bibinfo {author} {\bibfnamefont {D.}~\bibnamefont {Feng}}, \bibinfo {author} {\bibfnamefont {Z.}~\bibnamefont {Guan}}, \bibinfo {author} {\bibfnamefont {X.}~\bibnamefont {Wu}}, \bibinfo {author} {\bibfnamefont {Y.}~\bibnamefont {Wu}},\ and\ \bibinfo {author} {\bibfnamefont {C.}~\bibnamefont {Song}},\ }\bibfield  {title} {\bibinfo {title} {Classification of skyrmionic textures and extraction of {H}amiltonian parameters via machine learning},\ }\href {https://doi.org/10.1103/PhysRevApplied.21.034009} {\bibfield  {journal} {\bibinfo  {journal} {Phys. Rev. Appl.}\ }\textbf {\bibinfo {volume} {21}},\ \bibinfo {pages} {034009} (\bibinfo {year} {2024})}\BibitemShut {NoStop}%
\bibitem [{\citenamefont {Lupi}\ \emph {et~al.}()\citenamefont {Lupi}, \citenamefont {Fumega}, \citenamefont {Amini}, \citenamefont {Drost}, \citenamefont {Liljeroth},\ and\ \citenamefont {Lado}}]{lupi2026}%
  \BibitemOpen
  \bibfield  {author} {\bibinfo {author} {\bibfnamefont {G.}~\bibnamefont {Lupi}}, \bibinfo {author} {\bibfnamefont {A.~O.}\ \bibnamefont {Fumega}}, \bibinfo {author} {\bibfnamefont {M.}~\bibnamefont {Amini}}, \bibinfo {author} {\bibfnamefont {R.}~\bibnamefont {Drost}}, \bibinfo {author} {\bibfnamefont {P.}~\bibnamefont {Liljeroth}},\ and\ \bibinfo {author} {\bibfnamefont {J.~L.}\ \bibnamefont {Lado}},\ }\bibfield  {title} {\bibinfo {title} {Molecular hamiltonian learning from setpoint-dependent scanning tunneling spectroscopy},\ }\href {https://arxiv.org/abs/2601.19371} {\ }\Eprint {https://arxiv.org/abs/2601.19371} {arXiv:2601.19371} \BibitemShut {NoStop}%
\bibitem [{\citenamefont {Hernandes}\ \emph {et~al.}()\citenamefont {Hernandes}, \citenamefont {Rogers}, \citenamefont {Koch}, \citenamefont {Spriggs}, \citenamefont {Undseth}, \citenamefont {Chatterjee}, \citenamefont {Vandersypen},\ and\ \citenamefont {Greplova}}]{hernandes2026}%
  \BibitemOpen
  \bibfield  {author} {\bibinfo {author} {\bibfnamefont {V.}~\bibnamefont {Hernandes}}, \bibinfo {author} {\bibfnamefont {J.}~\bibnamefont {Rogers}}, \bibinfo {author} {\bibfnamefont {R.}~\bibnamefont {Koch}}, \bibinfo {author} {\bibfnamefont {T.}~\bibnamefont {Spriggs}}, \bibinfo {author} {\bibfnamefont {B.}~\bibnamefont {Undseth}}, \bibinfo {author} {\bibfnamefont {A.}~\bibnamefont {Chatterjee}}, \bibinfo {author} {\bibfnamefont {L.~M.~K.}\ \bibnamefont {Vandersypen}},\ and\ \bibinfo {author} {\bibfnamefont {E.}~\bibnamefont {Greplova}},\ }\bibfield  {title} {\bibinfo {title} {Reconstructing quantum dot charge stability diagrams with diffusion models},\ }\href {https://arxiv.org/abs/2603.26432} {\ }\Eprint {https://arxiv.org/abs/2603.26432} {arXiv:2603.26432} \BibitemShut {NoStop}%
\bibitem [{\citenamefont {Hofstadter}(1976)}]{Hofstadter1976}%
  \BibitemOpen
  \bibfield  {author} {\bibinfo {author} {\bibfnamefont {D.~R.}\ \bibnamefont {Hofstadter}},\ }\bibfield  {title} {\bibinfo {title} {Energy levels and wave functions of {B}loch electrons in rational and irrational magnetic fields},\ }\href {https://doi.org/10.1103/PhysRevB.14.2239} {\bibfield  {journal} {\bibinfo  {journal} {Phys. Rev. B}\ }\textbf {\bibinfo {volume} {14}},\ \bibinfo {pages} {2239} (\bibinfo {year} {1976})}\BibitemShut {NoStop}%
\bibitem [{\citenamefont {Wu}\ \emph {et~al.}(2018)\citenamefont {Wu}, \citenamefont {Lovorn}, \citenamefont {Tutuc},\ and\ \citenamefont {MacDonald}}]{Wu2018}%
  \BibitemOpen
  \bibfield  {author} {\bibinfo {author} {\bibfnamefont {F.}~\bibnamefont {Wu}}, \bibinfo {author} {\bibfnamefont {T.}~\bibnamefont {Lovorn}}, \bibinfo {author} {\bibfnamefont {E.}~\bibnamefont {Tutuc}},\ and\ \bibinfo {author} {\bibfnamefont {A.~H.}\ \bibnamefont {MacDonald}},\ }\bibfield  {title} {\bibinfo {title} {Hubbard model physics in transition metal dichalcogenide moir\'e bands},\ }\href {https://doi.org/10.1103/PhysRevLett.121.026402} {\bibfield  {journal} {\bibinfo  {journal} {Phys. Rev. Lett.}\ }\textbf {\bibinfo {volume} {121}},\ \bibinfo {pages} {026402} (\bibinfo {year} {2018})}\BibitemShut {NoStop}%
\bibitem [{\citenamefont {Regan}\ \emph {et~al.}(2020)\citenamefont {Regan}, \citenamefont {Wang}, \citenamefont {Jin}, \citenamefont {Bakti~Utama}, \citenamefont {Gao}, \citenamefont {Wei}, \citenamefont {Zhao}, \citenamefont {Zhao}, \citenamefont {Zhang}, \citenamefont {Yumigeta}, \citenamefont {Blei}, \citenamefont {Carlström}, \citenamefont {Watanabe}, \citenamefont {Taniguchi}, \citenamefont {Tongay}, \citenamefont {Crommie}, \citenamefont {Zettl},\ and\ \citenamefont {Wang}}]{regan2020}%
  \BibitemOpen
  \bibfield  {author} {\bibinfo {author} {\bibfnamefont {E.~C.}\ \bibnamefont {Regan}}, \bibinfo {author} {\bibfnamefont {D.}~\bibnamefont {Wang}}, \bibinfo {author} {\bibfnamefont {C.}~\bibnamefont {Jin}}, \bibinfo {author} {\bibfnamefont {M.~I.}\ \bibnamefont {Bakti~Utama}}, \bibinfo {author} {\bibfnamefont {B.}~\bibnamefont {Gao}}, \bibinfo {author} {\bibfnamefont {X.}~\bibnamefont {Wei}}, \bibinfo {author} {\bibfnamefont {S.}~\bibnamefont {Zhao}}, \bibinfo {author} {\bibfnamefont {W.}~\bibnamefont {Zhao}}, \bibinfo {author} {\bibfnamefont {Z.}~\bibnamefont {Zhang}}, \bibinfo {author} {\bibfnamefont {K.}~\bibnamefont {Yumigeta}}, \bibinfo {author} {\bibfnamefont {M.}~\bibnamefont {Blei}}, \bibinfo {author} {\bibfnamefont {J.~D.}\ \bibnamefont {Carlström}}, \bibinfo {author} {\bibfnamefont {K.}~\bibnamefont {Watanabe}}, \bibinfo {author} {\bibfnamefont {T.}~\bibnamefont {Taniguchi}}, \bibinfo {author} {\bibfnamefont {S.}~\bibnamefont {Tongay}}, \bibinfo {author} {\bibfnamefont {M.}~\bibnamefont {Crommie}},
  \bibinfo {author} {\bibfnamefont {A.}~\bibnamefont {Zettl}},\ and\ \bibinfo {author} {\bibfnamefont {F.}~\bibnamefont {Wang}},\ }\bibfield  {title} {\bibinfo {title} {Mott and generalized {W}igner crystal states in {WS}e$_2$/{WS}$_2$ moir{\'e} superlattices},\ }\href {https://doi.org/10.1038/s41586-020-2092-4} {\bibfield  {journal} {\bibinfo  {journal} {Nature}\ }\textbf {\bibinfo {volume} {579}},\ \bibinfo {pages} {359} (\bibinfo {year} {2020})}\BibitemShut {NoStop}%
\bibitem [{\citenamefont {Foutty}\ \emph {et~al.}(2025)\citenamefont {Foutty}, \citenamefont {Reddy}, \citenamefont {Kometter}, \citenamefont {Watanabe}, \citenamefont {Taniguchi}, \citenamefont {Devakul},\ and\ \citenamefont {Feldman}}]{foutty2025}%
  \BibitemOpen
  \bibfield  {author} {\bibinfo {author} {\bibfnamefont {B.~A.}\ \bibnamefont {Foutty}}, \bibinfo {author} {\bibfnamefont {A.~P.}\ \bibnamefont {Reddy}}, \bibinfo {author} {\bibfnamefont {C.~R.}\ \bibnamefont {Kometter}}, \bibinfo {author} {\bibfnamefont {K.}~\bibnamefont {Watanabe}}, \bibinfo {author} {\bibfnamefont {T.}~\bibnamefont {Taniguchi}}, \bibinfo {author} {\bibfnamefont {T.}~\bibnamefont {Devakul}},\ and\ \bibinfo {author} {\bibfnamefont {B.~E.}\ \bibnamefont {Feldman}},\ }\bibfield  {title} {\bibinfo {title} {Magnetic {H}ofstadter cascade in a twisted semiconductor homobilayer},\ }\href {https://doi.org/https://doi.org/10.1038/s41567-025-03083-5} {\bibfield  {journal} {\bibinfo  {journal} {Nat. Phys.}\ }\textbf {\bibinfo {volume} {21}},\ \bibinfo {pages} {1942} (\bibinfo {year} {2025})}\BibitemShut {NoStop}%
\bibitem [{\citenamefont {Kometter}\ \emph {et~al.}(2023)\citenamefont {Kometter}, \citenamefont {Yu}, \citenamefont {Devakul}, \citenamefont {Reddy}, \citenamefont {Zhang}, \citenamefont {Foutty}, \citenamefont {Watanabe}, \citenamefont {Taniguchi}, \citenamefont {Fu},\ and\ \citenamefont {Feldman}}]{kometter2023}%
  \BibitemOpen
  \bibfield  {author} {\bibinfo {author} {\bibfnamefont {C.~R.}\ \bibnamefont {Kometter}}, \bibinfo {author} {\bibfnamefont {J.}~\bibnamefont {Yu}}, \bibinfo {author} {\bibfnamefont {T.}~\bibnamefont {Devakul}}, \bibinfo {author} {\bibfnamefont {A.~P.}\ \bibnamefont {Reddy}}, \bibinfo {author} {\bibfnamefont {Y.}~\bibnamefont {Zhang}}, \bibinfo {author} {\bibfnamefont {B.~A.}\ \bibnamefont {Foutty}}, \bibinfo {author} {\bibfnamefont {K.}~\bibnamefont {Watanabe}}, \bibinfo {author} {\bibfnamefont {T.}~\bibnamefont {Taniguchi}}, \bibinfo {author} {\bibfnamefont {L.}~\bibnamefont {Fu}},\ and\ \bibinfo {author} {\bibfnamefont {B.~E.}\ \bibnamefont {Feldman}},\ }\bibfield  {title} {\bibinfo {title} {Hofstadter states and re-entrant charge order in a semiconductor moir{\'e} lattice},\ }\href {https://doi.org/10.1038/s41567-023-02195-0} {\bibfield  {journal} {\bibinfo  {journal} {Nat. Phys.}\ }\textbf {\bibinfo {volume} {19}},\ \bibinfo {pages} {1861} (\bibinfo {year} {2023})}\BibitemShut {NoStop}%
\bibitem [{\citenamefont {Zhao}\ \emph {et~al.}(2025)\citenamefont {Zhao}, \citenamefont {Wu}, \citenamefont {Ma}, \citenamefont {Liang}, \citenamefont {Lu}, \citenamefont {Gao},\ and\ \citenamefont {Xie}}]{Zhao2025}%
  \BibitemOpen
  \bibfield  {author} {\bibinfo {author} {\bibfnamefont {C.}~\bibnamefont {Zhao}}, \bibinfo {author} {\bibfnamefont {M.}~\bibnamefont {Wu}}, \bibinfo {author} {\bibfnamefont {Z.}~\bibnamefont {Ma}}, \bibinfo {author} {\bibfnamefont {M.}~\bibnamefont {Liang}}, \bibinfo {author} {\bibfnamefont {M.}~\bibnamefont {Lu}}, \bibinfo {author} {\bibfnamefont {J.-H.}\ \bibnamefont {Gao}},\ and\ \bibinfo {author} {\bibfnamefont {X.~C.}\ \bibnamefont {Xie}},\ }\bibfield  {title} {\bibinfo {title} {Hofstadter spectrum in a semiconductor moir\'e lattice},\ }\href {https://doi.org/10.1103/PhysRevB.111.L201302} {\bibfield  {journal} {\bibinfo  {journal} {Phys. Rev. B}\ }\textbf {\bibinfo {volume} {111}},\ \bibinfo {pages} {L201302} (\bibinfo {year} {2025})}\BibitemShut {NoStop}%
\bibitem [{\citenamefont {Foutty}\ \emph {et~al.}(2024)\citenamefont {Foutty}, \citenamefont {Kometter}, \citenamefont {Devakul}, \citenamefont {Reddy}, \citenamefont {Watanabe}, \citenamefont {Taniguchi}, \citenamefont {Fu},\ and\ \citenamefont {Feldman}}]{Foutty2024}%
  \BibitemOpen
  \bibfield  {author} {\bibinfo {author} {\bibfnamefont {B.~A.}\ \bibnamefont {Foutty}}, \bibinfo {author} {\bibfnamefont {C.~R.}\ \bibnamefont {Kometter}}, \bibinfo {author} {\bibfnamefont {T.}~\bibnamefont {Devakul}}, \bibinfo {author} {\bibfnamefont {A.~P.}\ \bibnamefont {Reddy}}, \bibinfo {author} {\bibfnamefont {K.}~\bibnamefont {Watanabe}}, \bibinfo {author} {\bibfnamefont {T.}~\bibnamefont {Taniguchi}}, \bibinfo {author} {\bibfnamefont {L.}~\bibnamefont {Fu}},\ and\ \bibinfo {author} {\bibfnamefont {B.~E.}\ \bibnamefont {Feldman}},\ }\bibfield  {title} {\bibinfo {title} {Mapping twist-tuned multiband topology in bilayer {WS}e$_2$},\ }\href {https://doi.org/10.1126/science.adi4728} {\bibfield  {journal} {\bibinfo  {journal} {Science}\ }\textbf {\bibinfo {volume} {384}},\ \bibinfo {pages} {343} (\bibinfo {year} {2024})}\BibitemShut {NoStop}%
\bibitem [{\citenamefont {Hunt}\ \emph {et~al.}(2013)\citenamefont {Hunt}, \citenamefont {Sanchez-Yamagishi}, \citenamefont {Young}, \citenamefont {Yankowitz}, \citenamefont {LeRoy}, \citenamefont {Watanabe}, \citenamefont {Taniguchi}, \citenamefont {Moon}, \citenamefont {Koshino}, \citenamefont {Jarillo-Herrero},\ and\ \citenamefont {Ashoori}}]{Hunt2013}%
  \BibitemOpen
  \bibfield  {author} {\bibinfo {author} {\bibfnamefont {B.}~\bibnamefont {Hunt}}, \bibinfo {author} {\bibfnamefont {J.~D.}\ \bibnamefont {Sanchez-Yamagishi}}, \bibinfo {author} {\bibfnamefont {A.~F.}\ \bibnamefont {Young}}, \bibinfo {author} {\bibfnamefont {M.}~\bibnamefont {Yankowitz}}, \bibinfo {author} {\bibfnamefont {B.~J.}\ \bibnamefont {LeRoy}}, \bibinfo {author} {\bibfnamefont {K.}~\bibnamefont {Watanabe}}, \bibinfo {author} {\bibfnamefont {T.}~\bibnamefont {Taniguchi}}, \bibinfo {author} {\bibfnamefont {P.}~\bibnamefont {Moon}}, \bibinfo {author} {\bibfnamefont {M.}~\bibnamefont {Koshino}}, \bibinfo {author} {\bibfnamefont {P.}~\bibnamefont {Jarillo-Herrero}},\ and\ \bibinfo {author} {\bibfnamefont {R.~C.}\ \bibnamefont {Ashoori}},\ }\bibfield  {title} {\bibinfo {title} {Massive {D}irac fermions and {H}ofstadter butterfly in a van der {W}aals heterostructure},\ }\href {https://doi.org/10.1126/science.1237240} {\bibfield  {journal} {\bibinfo  {journal} {Science}\ }\textbf {\bibinfo {volume} {340}},\
  \bibinfo {pages} {1427} (\bibinfo {year} {2013})}\BibitemShut {NoStop}%
\bibitem [{\citenamefont {Dean}\ \emph {et~al.}(2013)\citenamefont {Dean}, \citenamefont {Wang}, \citenamefont {Maher}, \citenamefont {Forsythe}, \citenamefont {Ghahari}, \citenamefont {Gao}, \citenamefont {Katoch}, \citenamefont {Ishigami}, \citenamefont {Moon}, \citenamefont {Koshino}, \citenamefont {Taniguchi}, \citenamefont {Watanabe}, \citenamefont {Shepard}, \citenamefont {Hone},\ and\ \citenamefont {Kim}}]{Dean2013}%
  \BibitemOpen
  \bibfield  {author} {\bibinfo {author} {\bibfnamefont {C.~R.}\ \bibnamefont {Dean}}, \bibinfo {author} {\bibfnamefont {L.}~\bibnamefont {Wang}}, \bibinfo {author} {\bibfnamefont {P.}~\bibnamefont {Maher}}, \bibinfo {author} {\bibfnamefont {C.}~\bibnamefont {Forsythe}}, \bibinfo {author} {\bibfnamefont {F.}~\bibnamefont {Ghahari}}, \bibinfo {author} {\bibfnamefont {Y.}~\bibnamefont {Gao}}, \bibinfo {author} {\bibfnamefont {J.}~\bibnamefont {Katoch}}, \bibinfo {author} {\bibfnamefont {M.}~\bibnamefont {Ishigami}}, \bibinfo {author} {\bibfnamefont {P.}~\bibnamefont {Moon}}, \bibinfo {author} {\bibfnamefont {M.}~\bibnamefont {Koshino}}, \bibinfo {author} {\bibfnamefont {T.}~\bibnamefont {Taniguchi}}, \bibinfo {author} {\bibfnamefont {K.}~\bibnamefont {Watanabe}}, \bibinfo {author} {\bibfnamefont {K.~L.}\ \bibnamefont {Shepard}}, \bibinfo {author} {\bibfnamefont {J.}~\bibnamefont {Hone}},\ and\ \bibinfo {author} {\bibfnamefont {P.}~\bibnamefont {Kim}},\ }\bibfield  {title} {\bibinfo {title} {{H}ofstadter’s
  butterfly and the fractal quantum {H}all effect in moir{\'e} superlattices},\ }\href {https://doi.org/10.1038/nature12186} {\bibfield  {journal} {\bibinfo  {journal} {Nature}\ }\textbf {\bibinfo {volume} {497}},\ \bibinfo {pages} {598} (\bibinfo {year} {2013})}\BibitemShut {NoStop}%
\bibitem [{\citenamefont {Ponomarenko}\ \emph {et~al.}(2013)\citenamefont {Ponomarenko}, \citenamefont {Gorbachev}, \citenamefont {Yu}, \citenamefont {Elias}, \citenamefont {Jalil}, \citenamefont {Patel}, \citenamefont {Mishchenko}, \citenamefont {Mayorov}, \citenamefont {Woods}, \citenamefont {Wallbank}, \citenamefont {Mucha-Kruczy{\'n}ski}, \citenamefont {Piot}, \citenamefont {Potemski}, \citenamefont {Grigorieva}, \citenamefont {Novoselov}, \citenamefont {Guinea}, \citenamefont {Fal’ko},\ and\ \citenamefont {Geim}}]{ponomarenko2013}%
  \BibitemOpen
  \bibfield  {author} {\bibinfo {author} {\bibfnamefont {L.~A.}\ \bibnamefont {Ponomarenko}}, \bibinfo {author} {\bibfnamefont {R.~V.}\ \bibnamefont {Gorbachev}}, \bibinfo {author} {\bibfnamefont {G.~L.}\ \bibnamefont {Yu}}, \bibinfo {author} {\bibfnamefont {D.~C.}\ \bibnamefont {Elias}}, \bibinfo {author} {\bibfnamefont {R.}~\bibnamefont {Jalil}}, \bibinfo {author} {\bibfnamefont {A.~A.}\ \bibnamefont {Patel}}, \bibinfo {author} {\bibfnamefont {A.}~\bibnamefont {Mishchenko}}, \bibinfo {author} {\bibfnamefont {A.~S.}\ \bibnamefont {Mayorov}}, \bibinfo {author} {\bibfnamefont {C.~R.}\ \bibnamefont {Woods}}, \bibinfo {author} {\bibfnamefont {J.~R.}\ \bibnamefont {Wallbank}}, \bibinfo {author} {\bibfnamefont {M.}~\bibnamefont {Mucha-Kruczy{\'n}ski}}, \bibinfo {author} {\bibfnamefont {B.~A.}\ \bibnamefont {Piot}}, \bibinfo {author} {\bibfnamefont {M.}~\bibnamefont {Potemski}}, \bibinfo {author} {\bibfnamefont {I.~V.}\ \bibnamefont {Grigorieva}}, \bibinfo {author} {\bibfnamefont {K.~S.}\ \bibnamefont {Novoselov}},
  \bibinfo {author} {\bibfnamefont {F.}~\bibnamefont {Guinea}}, \bibinfo {author} {\bibfnamefont {V.~I.}\ \bibnamefont {Fal’ko}},\ and\ \bibinfo {author} {\bibfnamefont {A.~K.}\ \bibnamefont {Geim}},\ }\bibfield  {title} {\bibinfo {title} {Cloning of {D}irac fermions in graphene superlattices},\ }\href {https://doi.org/10.1038/nature12187} {\bibfield  {journal} {\bibinfo  {journal} {Nature}\ }\textbf {\bibinfo {volume} {497}},\ \bibinfo {pages} {594} (\bibinfo {year} {2013})}\BibitemShut {NoStop}%
\bibitem [{\citenamefont {Mak}\ and\ \citenamefont {Shan}(2022)}]{mak2022}%
  \BibitemOpen
  \bibfield  {author} {\bibinfo {author} {\bibfnamefont {K.~F.}\ \bibnamefont {Mak}}\ and\ \bibinfo {author} {\bibfnamefont {J.}~\bibnamefont {Shan}},\ }\bibfield  {title} {\bibinfo {title} {Semiconductor moir{\'e} materials},\ }\href {https://doi.org/10.1038/s41565-022-01165-6} {\bibfield  {journal} {\bibinfo  {journal} {Nat. Nanotechnol.}\ }\textbf {\bibinfo {volume} {17}},\ \bibinfo {pages} {686} (\bibinfo {year} {2022})}\BibitemShut {NoStop}%
\bibitem [{\citenamefont {He}\ \emph {et~al.}(2021)\citenamefont {He}, \citenamefont {Zhou}, \citenamefont {Ye}, \citenamefont {Cho}, \citenamefont {Jeong}, \citenamefont {Meng},\ and\ \citenamefont {Wang}}]{He2021}%
  \BibitemOpen
  \bibfield  {author} {\bibinfo {author} {\bibfnamefont {F.}~\bibnamefont {He}}, \bibinfo {author} {\bibfnamefont {Y.}~\bibnamefont {Zhou}}, \bibinfo {author} {\bibfnamefont {Z.}~\bibnamefont {Ye}}, \bibinfo {author} {\bibfnamefont {S.-H.}\ \bibnamefont {Cho}}, \bibinfo {author} {\bibfnamefont {J.}~\bibnamefont {Jeong}}, \bibinfo {author} {\bibfnamefont {X.}~\bibnamefont {Meng}},\ and\ \bibinfo {author} {\bibfnamefont {Y.}~\bibnamefont {Wang}},\ }\bibfield  {title} {\bibinfo {title} {Moiré patterns in 2{D} materials: A review},\ }\href {https://doi.org/10.1021/acsnano.0c10435} {\bibfield  {journal} {\bibinfo  {journal} {ACS Nano}\ }\textbf {\bibinfo {volume} {15}},\ \bibinfo {pages} {5944} (\bibinfo {year} {2021})}\BibitemShut {NoStop}%
\bibitem [{\citenamefont {Lau}\ \emph {et~al.}(2022)\citenamefont {Lau}, \citenamefont {Bockrath}, \citenamefont {Mak},\ and\ \citenamefont {Zhang}}]{lau2022}%
  \BibitemOpen
  \bibfield  {author} {\bibinfo {author} {\bibfnamefont {C.~N.}\ \bibnamefont {Lau}}, \bibinfo {author} {\bibfnamefont {M.~W.}\ \bibnamefont {Bockrath}}, \bibinfo {author} {\bibfnamefont {K.~F.}\ \bibnamefont {Mak}},\ and\ \bibinfo {author} {\bibfnamefont {F.}~\bibnamefont {Zhang}},\ }\bibfield  {title} {\bibinfo {title} {Reproducibility in the fabrication and physics of moir{\'e} materials},\ }\href {https://doi.org/10.1038/s41586-021-04173-z} {\bibfield  {journal} {\bibinfo  {journal} {Nature}\ }\textbf {\bibinfo {volume} {602}},\ \bibinfo {pages} {41} (\bibinfo {year} {2022})}\BibitemShut {NoStop}%
\bibitem [{\citenamefont {Guo}\ \emph {et~al.}(2021)\citenamefont {Guo}, \citenamefont {Hu}, \citenamefont {Liu},\ and\ \citenamefont {Tian}}]{Hao-Wei2021}%
  \BibitemOpen
  \bibfield  {author} {\bibinfo {author} {\bibfnamefont {H.-W.}\ \bibnamefont {Guo}}, \bibinfo {author} {\bibfnamefont {Z.}~\bibnamefont {Hu}}, \bibinfo {author} {\bibfnamefont {Z.-B.}\ \bibnamefont {Liu}},\ and\ \bibinfo {author} {\bibfnamefont {J.-G.}\ \bibnamefont {Tian}},\ }\bibfield  {title} {\bibinfo {title} {Stacking of 2{D} materials},\ }\href {https://doi.org/https://doi.org/10.1002/adfm.202007810} {\bibfield  {journal} {\bibinfo  {journal} {Adv. Funct. Mater.}\ }\textbf {\bibinfo {volume} {31}},\ \bibinfo {pages} {2007810} (\bibinfo {year} {2021})}\BibitemShut {NoStop}%
\bibitem [{\citenamefont {Xu}\ \emph {et~al.}(2016)\citenamefont {Xu}, \citenamefont {Wu}, \citenamefont {Lu}, \citenamefont {Han}, \citenamefont {Long}, \citenamefont {Chen}, \citenamefont {Han}, \citenamefont {Ye}, \citenamefont {Wu}, \citenamefont {Lin}, \citenamefont {Shen}, \citenamefont {Cai}, \citenamefont {He}, \citenamefont {Zhang}, \citenamefont {Lortz}, \citenamefont {Cheng},\ and\ \citenamefont {Wang}}]{Xu2016}%
  \BibitemOpen
  \bibfield  {author} {\bibinfo {author} {\bibfnamefont {S.}~\bibnamefont {Xu}}, \bibinfo {author} {\bibfnamefont {Z.}~\bibnamefont {Wu}}, \bibinfo {author} {\bibfnamefont {H.}~\bibnamefont {Lu}}, \bibinfo {author} {\bibfnamefont {Y.}~\bibnamefont {Han}}, \bibinfo {author} {\bibfnamefont {G.}~\bibnamefont {Long}}, \bibinfo {author} {\bibfnamefont {X.}~\bibnamefont {Chen}}, \bibinfo {author} {\bibfnamefont {T.}~\bibnamefont {Han}}, \bibinfo {author} {\bibfnamefont {W.}~\bibnamefont {Ye}}, \bibinfo {author} {\bibfnamefont {Y.}~\bibnamefont {Wu}}, \bibinfo {author} {\bibfnamefont {J.}~\bibnamefont {Lin}}, \bibinfo {author} {\bibfnamefont {J.}~\bibnamefont {Shen}}, \bibinfo {author} {\bibfnamefont {Y.}~\bibnamefont {Cai}}, \bibinfo {author} {\bibfnamefont {Y.}~\bibnamefont {He}}, \bibinfo {author} {\bibfnamefont {F.}~\bibnamefont {Zhang}}, \bibinfo {author} {\bibfnamefont {R.}~\bibnamefont {Lortz}}, \bibinfo {author} {\bibfnamefont {C.}~\bibnamefont {Cheng}},\ and\ \bibinfo {author} {\bibfnamefont
  {N.}~\bibnamefont {Wang}},\ }\bibfield  {title} {\bibinfo {title} {Universal low-temperature {O}hmic contacts for quantum transport in transition metal dichalcogenides},\ }\href {https://doi.org/10.1088/2053-1583/3/2/021007} {\bibfield  {journal} {\bibinfo  {journal} {2D Mater.}\ }\textbf {\bibinfo {volume} {3}},\ \bibinfo {pages} {021007} (\bibinfo {year} {2016})}\BibitemShut {NoStop}%
\bibitem [{\citenamefont {Jain}\ \emph {et~al.}(2019)\citenamefont {Jain}, \citenamefont {Szab{\'o}}, \citenamefont {Parzefall}, \citenamefont {Bonvin}, \citenamefont {Taniguchi}, \citenamefont {Watanabe}, \citenamefont {Bharadwaj}, \citenamefont {Luisier},\ and\ \citenamefont {Novotny}}]{jain2019}%
  \BibitemOpen
  \bibfield  {author} {\bibinfo {author} {\bibfnamefont {A.}~\bibnamefont {Jain}}, \bibinfo {author} {\bibfnamefont {{\'A}.}~\bibnamefont {Szab{\'o}}}, \bibinfo {author} {\bibfnamefont {M.}~\bibnamefont {Parzefall}}, \bibinfo {author} {\bibfnamefont {E.}~\bibnamefont {Bonvin}}, \bibinfo {author} {\bibfnamefont {T.}~\bibnamefont {Taniguchi}}, \bibinfo {author} {\bibfnamefont {K.}~\bibnamefont {Watanabe}}, \bibinfo {author} {\bibfnamefont {P.}~\bibnamefont {Bharadwaj}}, \bibinfo {author} {\bibfnamefont {M.}~\bibnamefont {Luisier}},\ and\ \bibinfo {author} {\bibfnamefont {L.}~\bibnamefont {Novotny}},\ }\bibfield  {title} {\bibinfo {title} {One-dimensional edge contacts to a monolayer semiconductor},\ }\href {https://doi.org/doi: 10.1021/acs.nanolett.9b02166} {\bibfield  {journal} {\bibinfo  {journal} {Nano Lett.}\ }\textbf {\bibinfo {volume} {19}},\ \bibinfo {pages} {6914} (\bibinfo {year} {2019})}\BibitemShut {NoStop}%
\bibitem [{\citenamefont {Bruus}\ and\ \citenamefont {Flensberg}(2004)}]{bruus2004many}%
  \BibitemOpen
  \bibfield  {author} {\bibinfo {author} {\bibfnamefont {H.}~\bibnamefont {Bruus}}\ and\ \bibinfo {author} {\bibfnamefont {K.}~\bibnamefont {Flensberg}},\ }\href@noop {} {\emph {\bibinfo {title} {Many-body quantum theory in condensed matter physics: {A}n introduction}}}\ (\bibinfo  {publisher} {Oxford University Press},\ \bibinfo {address} {Oxford},\ \bibinfo {year} {2004})\BibitemShut {NoStop}%
\bibitem [{pyq()}]{pyqula}%
  \BibitemOpen
  \href@noop {} {\bibinfo {title} {P{Y}{Q}{U}{L}{A} {L}ibrary}},\ \bibinfo {howpublished} {\url{https://github.com/joselado/pyqula}}\BibitemShut {NoStop}%
\bibitem [{\citenamefont {Klitzing}\ \emph {et~al.}(1980)\citenamefont {Klitzing}, \citenamefont {Dorda},\ and\ \citenamefont {Pepper}}]{Klitzing1980}%
  \BibitemOpen
  \bibfield  {author} {\bibinfo {author} {\bibfnamefont {K.~v.}\ \bibnamefont {Klitzing}}, \bibinfo {author} {\bibfnamefont {G.}~\bibnamefont {Dorda}},\ and\ \bibinfo {author} {\bibfnamefont {M.}~\bibnamefont {Pepper}},\ }\bibfield  {title} {\bibinfo {title} {New method for high-accuracy determination of the fine-structure constant based on quantized {H}all resistance},\ }\href {https://doi.org/10.1103/PhysRevLett.45.494} {\bibfield  {journal} {\bibinfo  {journal} {Phys. Rev. Lett.}\ }\textbf {\bibinfo {volume} {45}},\ \bibinfo {pages} {494} (\bibinfo {year} {1980})}\BibitemShut {NoStop}%
\bibitem [{\citenamefont {Halperin}(1982)}]{Halperin1982}%
  \BibitemOpen
  \bibfield  {author} {\bibinfo {author} {\bibfnamefont {B.~I.}\ \bibnamefont {Halperin}},\ }\bibfield  {title} {\bibinfo {title} {Quantized {H}all conductance, current-carrying edge states, and the existence of extended states in a two-dimensional disordered potential},\ }\href {https://doi.org/10.1103/PhysRevB.25.2185} {\bibfield  {journal} {\bibinfo  {journal} {Phys. Rev. B}\ }\textbf {\bibinfo {volume} {25}},\ \bibinfo {pages} {2185} (\bibinfo {year} {1982})}\BibitemShut {NoStop}%
\bibitem [{\citenamefont {Khosravian}\ \emph {et~al.}(2024)\citenamefont {Khosravian}, \citenamefont {Koch},\ and\ \citenamefont {Lado}}]{Khosravian2024}%
  \BibitemOpen
  \bibfield  {author} {\bibinfo {author} {\bibfnamefont {M.}~\bibnamefont {Khosravian}}, \bibinfo {author} {\bibfnamefont {R.}~\bibnamefont {Koch}},\ and\ \bibinfo {author} {\bibfnamefont {J.~L.}\ \bibnamefont {Lado}},\ }\bibfield  {title} {\bibinfo {title} {Hamiltonian learning with real-space impurity tomography in topological moiré superconductors},\ }\href {https://doi.org/10.1088/2515-7639/ad1c04} {\bibfield  {journal} {\bibinfo  {journal} {J. Phys.: Mater.}\ }\textbf {\bibinfo {volume} {7}},\ \bibinfo {pages} {015012} (\bibinfo {year} {2024})}\BibitemShut {NoStop}%
\bibitem [{\citenamefont {Tang}\ \emph {et~al.}(2020)\citenamefont {Tang}, \citenamefont {Li}, \citenamefont {Li}, \citenamefont {Xu}, \citenamefont {Liu}, \citenamefont {Barmak}, \citenamefont {Watanabe}, \citenamefont {Taniguchi}, \citenamefont {MacDonald}, \citenamefont {Shan},\ and\ \citenamefont {Mak}}]{tang2020}%
  \BibitemOpen
  \bibfield  {author} {\bibinfo {author} {\bibfnamefont {Y.}~\bibnamefont {Tang}}, \bibinfo {author} {\bibfnamefont {L.}~\bibnamefont {Li}}, \bibinfo {author} {\bibfnamefont {T.}~\bibnamefont {Li}}, \bibinfo {author} {\bibfnamefont {Y.}~\bibnamefont {Xu}}, \bibinfo {author} {\bibfnamefont {S.}~\bibnamefont {Liu}}, \bibinfo {author} {\bibfnamefont {K.}~\bibnamefont {Barmak}}, \bibinfo {author} {\bibfnamefont {K.}~\bibnamefont {Watanabe}}, \bibinfo {author} {\bibfnamefont {T.}~\bibnamefont {Taniguchi}}, \bibinfo {author} {\bibfnamefont {A.~H.}\ \bibnamefont {MacDonald}}, \bibinfo {author} {\bibfnamefont {J.}~\bibnamefont {Shan}},\ and\ \bibinfo {author} {\bibfnamefont {K.~F.}\ \bibnamefont {Mak}},\ }\bibfield  {title} {\bibinfo {title} {Simulation of {H}ubbard model physics in {WS}e$_2$/{WS}$_2$ moir{\'e} superlattices},\ }\href {https://doi.org/10.1038/s41586-020-2085-3} {\bibfield  {journal} {\bibinfo  {journal} {Nature}\ }\textbf {\bibinfo {volume} {579}},\ \bibinfo {pages} {353} (\bibinfo {year}
  {2020})}\BibitemShut {NoStop}%
\bibitem [{\citenamefont {Campbell}\ \emph {et~al.}(2022)\citenamefont {Campbell}, \citenamefont {Brotons-Gisbert}, \citenamefont {Baek}, \citenamefont {Vitale}, \citenamefont {Taniguchi}, \citenamefont {Watanabe}, \citenamefont {Lischner},\ and\ \citenamefont {Gerardot}}]{campbell2022}%
  \BibitemOpen
  \bibfield  {author} {\bibinfo {author} {\bibfnamefont {A.~J.}\ \bibnamefont {Campbell}}, \bibinfo {author} {\bibfnamefont {M.}~\bibnamefont {Brotons-Gisbert}}, \bibinfo {author} {\bibfnamefont {H.}~\bibnamefont {Baek}}, \bibinfo {author} {\bibfnamefont {V.}~\bibnamefont {Vitale}}, \bibinfo {author} {\bibfnamefont {T.}~\bibnamefont {Taniguchi}}, \bibinfo {author} {\bibfnamefont {K.}~\bibnamefont {Watanabe}}, \bibinfo {author} {\bibfnamefont {J.}~\bibnamefont {Lischner}},\ and\ \bibinfo {author} {\bibfnamefont {B.~D.}\ \bibnamefont {Gerardot}},\ }\bibfield  {title} {\bibinfo {title} {Exciton-polarons in the presence of strongly correlated electronic states in a {M}o{S}e$_2$/{WS}e$_2$ moir{\'e} superlattice},\ }\href {https://doi.org/10.1038/s41699-022-00358-w} {\bibfield  {journal} {\bibinfo  {journal} {npj 2D Mater. Appl.}\ }\textbf {\bibinfo {volume} {6}},\ \bibinfo {pages} {79} (\bibinfo {year} {2022})}\BibitemShut {NoStop}%
\bibitem [{\citenamefont {Regan}\ \emph {et~al.}(2024)\citenamefont {Regan}, \citenamefont {Lu}, \citenamefont {Wang}, \citenamefont {Zhang}, \citenamefont {Devakul}, \citenamefont {Nie}, \citenamefont {Zhang}, \citenamefont {Zhao}, \citenamefont {Watanabe}, \citenamefont {Taniguchi}, \citenamefont {Tongay}, \citenamefont {Crommie}, \citenamefont {Zettl},\ and\ \citenamefont {Wang}}]{regan2024}%
  \BibitemOpen
  \bibfield  {author} {\bibinfo {author} {\bibfnamefont {E.~C.}\ \bibnamefont {Regan}}, \bibinfo {author} {\bibfnamefont {Z.}~\bibnamefont {Lu}}, \bibinfo {author} {\bibfnamefont {D.}~\bibnamefont {Wang}}, \bibinfo {author} {\bibfnamefont {Y.}~\bibnamefont {Zhang}}, \bibinfo {author} {\bibfnamefont {T.}~\bibnamefont {Devakul}}, \bibinfo {author} {\bibfnamefont {J.~H.}\ \bibnamefont {Nie}}, \bibinfo {author} {\bibfnamefont {Z.}~\bibnamefont {Zhang}}, \bibinfo {author} {\bibfnamefont {W.}~\bibnamefont {Zhao}}, \bibinfo {author} {\bibfnamefont {K.}~\bibnamefont {Watanabe}}, \bibinfo {author} {\bibfnamefont {T.}~\bibnamefont {Taniguchi}}, \bibinfo {author} {\bibfnamefont {S.}~\bibnamefont {Tongay}}, \bibinfo {author} {\bibfnamefont {M.}~\bibnamefont {Crommie}}, \bibinfo {author} {\bibfnamefont {A.}~\bibnamefont {Zettl}},\ and\ \bibinfo {author} {\bibfnamefont {F.}~\bibnamefont {Wang}},\ }\bibfield  {title} {\bibinfo {title} {Spin transport of a doped {M}ott insulator in moir{\'e} heterostructures},\ }\href
  {https://doi.org/10.1038/s41467-024-54633-z} {\bibfield  {journal} {\bibinfo  {journal} {Nat. Commun.}\ }\textbf {\bibinfo {volume} {15}},\ \bibinfo {pages} {10252} (\bibinfo {year} {2024})}\BibitemShut {NoStop}%
\bibitem [{\citenamefont {Kiese}\ \emph {et~al.}(2022)\citenamefont {Kiese}, \citenamefont {He}, \citenamefont {Hickey}, \citenamefont {Rubio},\ and\ \citenamefont {Kennes}}]{kiese2022}%
  \BibitemOpen
  \bibfield  {author} {\bibinfo {author} {\bibfnamefont {D.}~\bibnamefont {Kiese}}, \bibinfo {author} {\bibfnamefont {Y.}~\bibnamefont {He}}, \bibinfo {author} {\bibfnamefont {C.}~\bibnamefont {Hickey}}, \bibinfo {author} {\bibfnamefont {A.}~\bibnamefont {Rubio}},\ and\ \bibinfo {author} {\bibfnamefont {D.~M.}\ \bibnamefont {Kennes}},\ }\bibfield  {title} {\bibinfo {title} {{TMD}s as a platform for spin liquid physics: A strong coupling study of twisted bilayer {WS}e$_2$},\ }\href {https://doi.org/10.1063/5.0077901} {\bibfield  {journal} {\bibinfo  {journal} {APL Mater.}\ }\textbf {\bibinfo {volume} {10}},\ \bibinfo {pages} {031113} (\bibinfo {year} {2022})}\BibitemShut {NoStop}%
\bibitem [{\citenamefont {Lee}\ \emph {et~al.}(2018)\citenamefont {Lee}, \citenamefont {Richardella}, \citenamefont {Fraleigh}, \citenamefont {Liu}, \citenamefont {Zhao},\ and\ \citenamefont {Samarth}}]{lee2018}%
  \BibitemOpen
  \bibfield  {author} {\bibinfo {author} {\bibfnamefont {J.~S.}\ \bibnamefont {Lee}}, \bibinfo {author} {\bibfnamefont {A.}~\bibnamefont {Richardella}}, \bibinfo {author} {\bibfnamefont {R.~D.}\ \bibnamefont {Fraleigh}}, \bibinfo {author} {\bibfnamefont {C.-x.}\ \bibnamefont {Liu}}, \bibinfo {author} {\bibfnamefont {W.}~\bibnamefont {Zhao}},\ and\ \bibinfo {author} {\bibfnamefont {N.}~\bibnamefont {Samarth}},\ }\bibfield  {title} {\bibinfo {title} {Engineering the breaking of time-reversal symmetry in gate-tunable hybrid ferromagnet/topological insulator heterostructures},\ }\href {https://doi.org/10.1038/s41535-018-0123-2} {\bibfield  {journal} {\bibinfo  {journal} {npj Quantum Mater.}\ }\textbf {\bibinfo {volume} {3}},\ \bibinfo {pages} {51} (\bibinfo {year} {2018})}\BibitemShut {NoStop}%
\bibitem [{\citenamefont {König}\ \emph {et~al.}(2007)\citenamefont {König}, \citenamefont {Wiedmann}, \citenamefont {Brune}, \citenamefont {Roth}, \citenamefont {Buhmann}, \citenamefont {Molenkamp}, \citenamefont {Qi},\ and\ \citenamefont {Zhang}}]{konig2007}%
  \BibitemOpen
  \bibfield  {author} {\bibinfo {author} {\bibfnamefont {M.}~\bibnamefont {König}}, \bibinfo {author} {\bibfnamefont {S.}~\bibnamefont {Wiedmann}}, \bibinfo {author} {\bibfnamefont {C.}~\bibnamefont {Brune}}, \bibinfo {author} {\bibfnamefont {A.}~\bibnamefont {Roth}}, \bibinfo {author} {\bibfnamefont {H.}~\bibnamefont {Buhmann}}, \bibinfo {author} {\bibfnamefont {L.~W.}\ \bibnamefont {Molenkamp}}, \bibinfo {author} {\bibfnamefont {X.-L.}\ \bibnamefont {Qi}},\ and\ \bibinfo {author} {\bibfnamefont {S.-C.}\ \bibnamefont {Zhang}},\ }\bibfield  {title} {\bibinfo {title} {Quantum spin {H}all insulator state in {H}g{T}e quantum wells},\ }\href {https://doi.org/10.1126/science.1148047} {\bibfield  {journal} {\bibinfo  {journal} {Science}\ }\textbf {\bibinfo {volume} {318}},\ \bibinfo {pages} {766} (\bibinfo {year} {2007})}\BibitemShut {NoStop}%
\bibitem [{\citenamefont {Mukhopadhyay}\ \emph {et~al.}(2021)\citenamefont {Mukhopadhyay}, \citenamefont {Kanungo},\ and\ \citenamefont {Rahaman}}]{mukhopadhyay2021}%
  \BibitemOpen
  \bibfield  {author} {\bibinfo {author} {\bibfnamefont {A.}~\bibnamefont {Mukhopadhyay}}, \bibinfo {author} {\bibfnamefont {S.}~\bibnamefont {Kanungo}},\ and\ \bibinfo {author} {\bibfnamefont {H.}~\bibnamefont {Rahaman}},\ }\bibfield  {title} {\bibinfo {title} {The effect of the stacking arrangement on the device behavior of bilayer {M}o{S}$_2$ {FET}s},\ }\href {https://doi.org/10.1007/s10825-020-01636-w} {\bibfield  {journal} {\bibinfo  {journal} {J. Comput. Electron.}\ }\textbf {\bibinfo {volume} {20}},\ \bibinfo {pages} {161} (\bibinfo {year} {2021})}\BibitemShut {NoStop}%
\bibitem [{\citenamefont {Schmidt}\ \emph {et~al.}(2014)\citenamefont {Schmidt}, \citenamefont {Rode}, \citenamefont {Smirnov},\ and\ \citenamefont {Haug}}]{schmidt2014}%
  \BibitemOpen
  \bibfield  {author} {\bibinfo {author} {\bibfnamefont {H.}~\bibnamefont {Schmidt}}, \bibinfo {author} {\bibfnamefont {J.~C.}\ \bibnamefont {Rode}}, \bibinfo {author} {\bibfnamefont {D.}~\bibnamefont {Smirnov}},\ and\ \bibinfo {author} {\bibfnamefont {R.~J.}\ \bibnamefont {Haug}},\ }\bibfield  {title} {\bibinfo {title} {Superlattice structures in twisted bilayers of folded graphene},\ }\href {https://doi.org/10.1038/ncomms6742} {\bibfield  {journal} {\bibinfo  {journal} {Nat. Commun.}\ }\textbf {\bibinfo {volume} {5}},\ \bibinfo {pages} {5742} (\bibinfo {year} {2014})}\BibitemShut {NoStop}%
\bibitem [{\citenamefont {Talkington}\ \emph {et~al.}()\citenamefont {Talkington}, \citenamefont {Mallick}, \citenamefont {Chen}, \citenamefont {Mead}, \citenamefont {Yang}, \citenamefont {Kim}, \citenamefont {Adam}, \citenamefont {Wu}, \citenamefont {Brahlek},\ and\ \citenamefont {Mele}}]{talkington2025}%
  \BibitemOpen
  \bibfield  {author} {\bibinfo {author} {\bibfnamefont {S.}~\bibnamefont {Talkington}}, \bibinfo {author} {\bibfnamefont {D.}~\bibnamefont {Mallick}}, \bibinfo {author} {\bibfnamefont {A.-H.}\ \bibnamefont {Chen}}, \bibinfo {author} {\bibfnamefont {B.~F.}\ \bibnamefont {Mead}}, \bibinfo {author} {\bibfnamefont {S.-J.}\ \bibnamefont {Yang}}, \bibinfo {author} {\bibfnamefont {C.-J.}\ \bibnamefont {Kim}}, \bibinfo {author} {\bibfnamefont {S.}~\bibnamefont {Adam}}, \bibinfo {author} {\bibfnamefont {L.}~\bibnamefont {Wu}}, \bibinfo {author} {\bibfnamefont {M.}~\bibnamefont {Brahlek}},\ and\ \bibinfo {author} {\bibfnamefont {E.~J.}\ \bibnamefont {Mele}},\ }\bibfield  {title} {\bibinfo {title} {Weak localization and universal conductance fluctuations in large area twisted bilayer graphene},\ }\href {https://arxiv.org/abs/2511.07334} {\ }\Eprint {https://arxiv.org/abs/2511.07334} {arXiv:2511.07334} \BibitemShut {NoStop}%
\bibitem [{\citenamefont {Guerrero}\ \emph {et~al.}(2025)\citenamefont {Guerrero}, \citenamefont {Nguyen}, \citenamefont {Romeral}, \citenamefont {Cummings}, \citenamefont {Garcia}, \citenamefont {Charlier},\ and\ \citenamefont {Roche}}]{Guerrero2025}%
  \BibitemOpen
  \bibfield  {author} {\bibinfo {author} {\bibfnamefont {P.~A.}\ \bibnamefont {Guerrero}}, \bibinfo {author} {\bibfnamefont {V.-H.}\ \bibnamefont {Nguyen}}, \bibinfo {author} {\bibfnamefont {J.~M.}\ \bibnamefont {Romeral}}, \bibinfo {author} {\bibfnamefont {A.~W.}\ \bibnamefont {Cummings}}, \bibinfo {author} {\bibfnamefont {J.-H.}\ \bibnamefont {Garcia}}, \bibinfo {author} {\bibfnamefont {J.-C.}\ \bibnamefont {Charlier}},\ and\ \bibinfo {author} {\bibfnamefont {S.}~\bibnamefont {Roche}},\ }\bibfield  {title} {\bibinfo {title} {Disorder-induced delocalization in magic-angle twisted bilayer graphene},\ }\href {https://doi.org/10.1103/PhysRevLett.134.126301} {\bibfield  {journal} {\bibinfo  {journal} {Phys. Rev. Lett.}\ }\textbf {\bibinfo {volume} {134}},\ \bibinfo {pages} {126301} (\bibinfo {year} {2025})}\BibitemShut {NoStop}%
\end{thebibliography}%

\end{document}